\documentclass[sigconf]{acmart}

\AtBeginDocument{%
  }
    
\copyrightyear{2025}
\acmYear{2025}
\acmConference[SC '25]{The International Conference for High Performance
Computing, Networking, Storage and Analysis}{November 16--21, 2025}{St
Louis, MO, USA}
\acmBooktitle{The International Conference for High Performance Computing,
Networking, Storage and Analysis (SC '25), November 16--21, 2025, St Louis,
MO, USA}
\acmDOI{10.1145/3712285.3759777}
\acmISBN{979-8-4007-1466-5/2025/11}

\usepackage[]{hyperref}
\usepackage[noend]{algorithm}
\usepackage[noend]{algpseudocode}
\usepackage{adjustbox}
\usepackage{graphicx}
\usepackage{subcaption}
\usepackage{caption}
\usepackage{float} 
\usepackage{longtable}
\usepackage{multirow}
\usepackage{amsmath} 
\usepackage{tikz}

\newcommand*\circled[1]{\tikz[baseline=(char.base)]{
            \node[shape=circle,draw,inner sep=0.6pt] (char) {#1};}}

\newcommand{\sysname}{UpANNS}

\begin{document}


\title{UpANNS: Enhancing Billion-Scale ANNS Efficiency with Real-World PIM Architecture}



\author{Sitian Chen}
\affiliation{%
  \institution{Hong Kong Baptist University}
  \country{}
}
\author{Amelie Chi Zhou}
\authornote{Corresponding author.}
\affiliation{%
  \institution{Hong Kong Baptist University}
  \country{}
}
\author{Yucheng Shi}
\affiliation{%
  \institution{Hong Kong Baptist University}
  \country{}
}


\author{Yusen Li}
\affiliation{%
  \institution{Nankai University}
  \country{}}

\author{Xin Yao}
\affiliation{%
  \institution{Huawei}
  \country{}
}

\renewcommand{\shortauthors}{Chen et al.}

\begin{abstract}


Approximate Nearest Neighbor Search (ANNS) is a critical component of modern AI systems, such as recommendation engines and retrieval-augmented large language models (RAG-LLMs). However, scaling ANNS to billion-entry datasets exposes critical inefficiencies: CPU-based solutions are bottlenecked by memory bandwidth limitations, while GPU implementations underutilize hardware resources, leading to suboptimal performance and energy consumption.
To address these challenges, we introduce \emph{UpANNS}, a novel framework leveraging Processing-in-Memory (PIM) architecture to accelerate billion-scale ANNS.
UpANNS integrates four key innovations, including 1) architecture-aware data placement to minimize latency through workload balancing, 2) dynamic resource management for optimal PIM utilization, 3) co-occurrence optimized encoding to reduce redundant computations, and 4) an early-pruning strategy for efficient top-k selection.
Evaluation on commercial UPMEM hardware demonstrates that UpANNS achieves 4.3x higher QPS than CPU-based Faiss, while matching GPU performance with 2.3x greater energy efficiency.
Its near-linear scalability ensures practicality for growing datasets, making it ideal for applications like real-time LLM serving and large-scale retrieval systems.

\end{abstract}

\maketitle



\section{Introduction}\label{sec:intro}


Approximate Nearest Neighbor Search (ANNS) is a foundational technology for modern AI systems, enabling applications like real-time recommendations~\cite{li2024ndrec,wang2023ems,chen2022approximate} and retrieval-augmented large language models (RAG-LLMs)~\cite{shi2023context,khandelwal2019generalization, jiang2023chameleon}.
As datasets grow to billions of entries, which is common in social networks~\cite{pmlr-v235-zhai24a} or enterprise-scale LLM serving, efficient ANNS becomes critical. 

ANNS algorithms can be categorized into four classes: hash-based~\cite{andoni2015practical, zheng2016lazylsh, gong2020idec}, tree-based~\cite{chen2023parallelnn, zheng2020pm, arora2018hd}, graph-based~\cite{9726805, DBLP:conf/ppopp/ManoharSBD0S024, wang2024ndsearch} and compression-based~\cite{johnson2019billion, juno-asplos24, vectorsearch-sc23}. 
While graph-based, tree-based and hash-based methods excel at million-scale searches, they encounter scalability challenges for billion-scale datasets due to prohibitive memory requirements. For example, HNSW~\cite{hnsw}, a state-of-the-art graph-based method, requires 60-450 bytes of memory per vertex. This results in a memory requirement of up to 450GB for a billion-vertex graph, making it impractical for real-world deployments.
In contrast, compression-based methods dramatically reduce storage footprints by encoding data into compact representations, enabling scalable and cost-efficient billion-scale searches. 
Thus, this paper focuses mainly on compression-based ANNS algorithms. Specifically, we focus on IVFPQ~\cite{jegou2010product}, one of the most popular compression-based algorithms and widely used for accelerating massive industrial systems (e.g., recommendation systems in ByteDance~\cite{bytedance-ivfpq} and long-context LLM serving~\cite{zhang2025pqcacheproductquantizationbasedkvcache}).

\begin{figure}[t]
    \centering
    \begin{subfigure}[b]{0.23\textwidth}
        \centering
        \includegraphics[width=\textwidth]{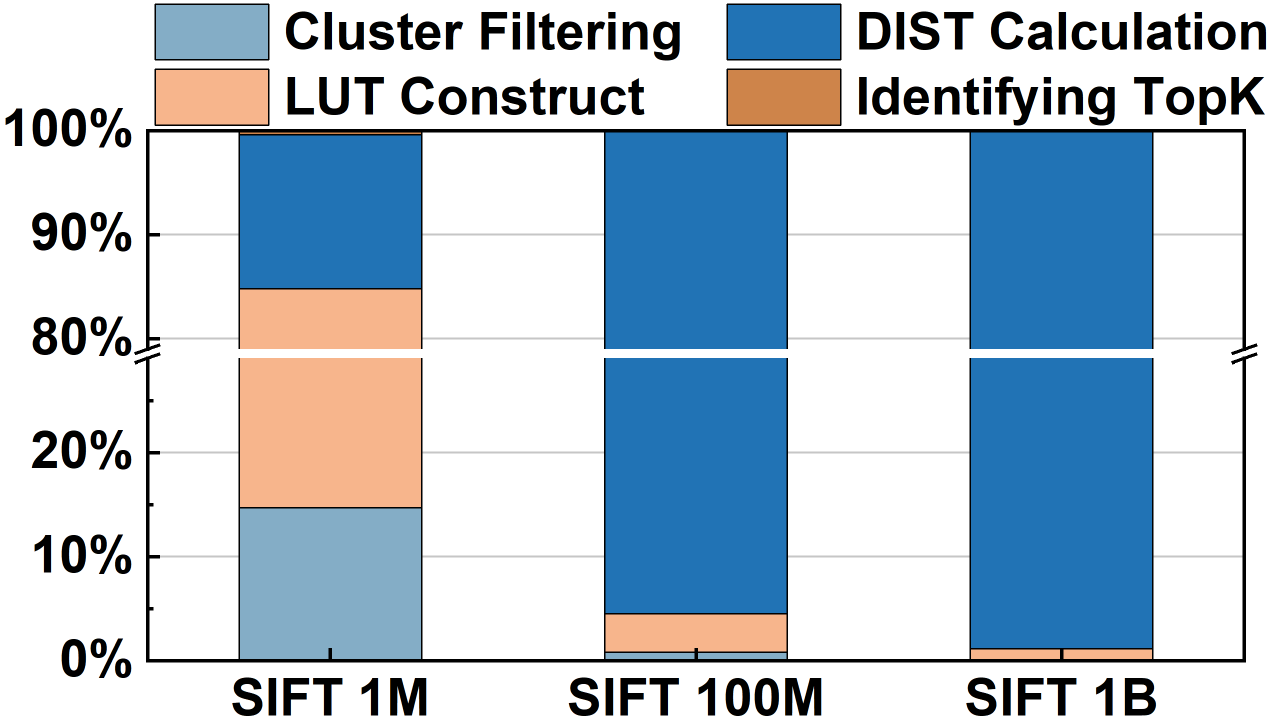}
        \caption{CPU}
        \label{fig:breakdown:cpu}
    \end{subfigure}
    \hfil
    \begin{subfigure}[b]{0.23\textwidth}
        \centering
        \includegraphics[width=\textwidth]{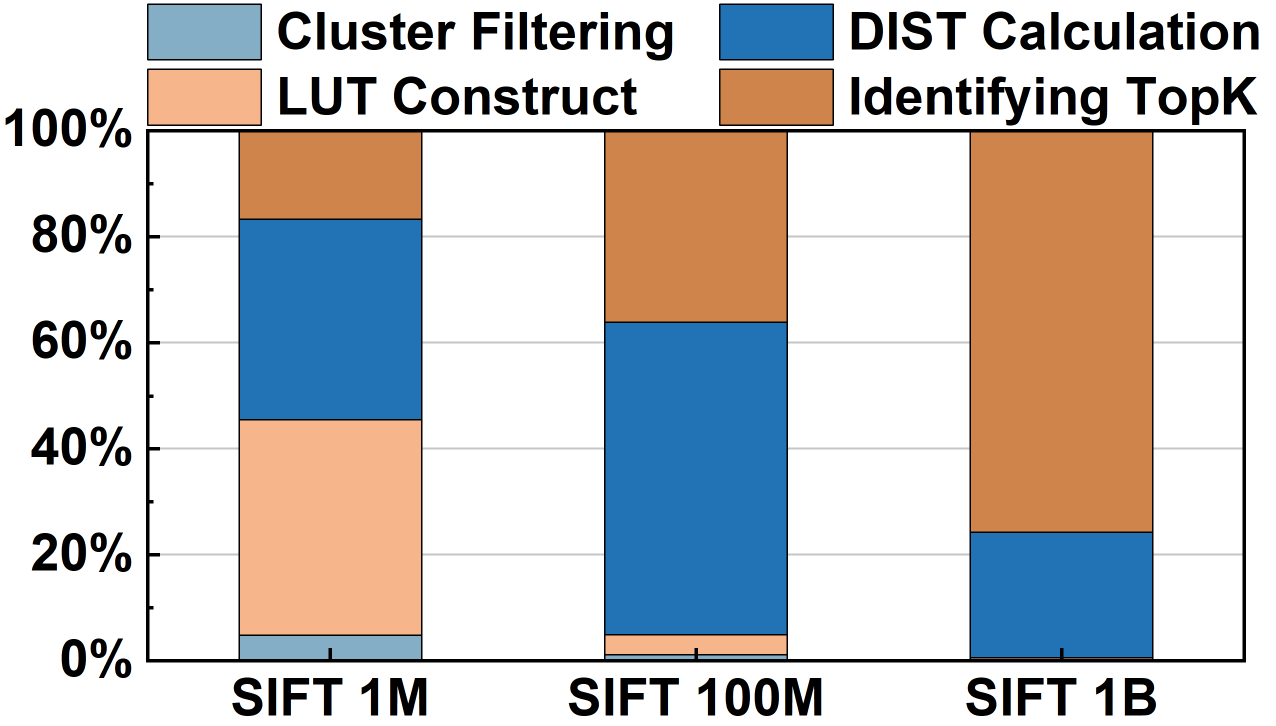}
        \caption{GPU}
        \label{fig:breakdown:gpu}
    \end{subfigure}
    \caption{Query processing time breakdown of the IVFPQ~\cite{jegou2010product} algorithm, using SIFT dataset~\cite{jegou2011searching} at different scales (1M, 100M and 1B). Hardware specifications see Section~\ref{sec::exp}.}
    \label{fig:breakdown}
\end{figure}

\textbf{Limitation of Existing Architectures.}
IVFPQ integrates two key techniques, including cluster-based filtering (IVF) and product quantization (PQ). IVF partitions data into clusters to narrow search scope and PQ compresses data into compact codes, greatly reducing the storage requirements while maintaining search accuracy. However, our analysis of IVFPQ's four-stage pipeline, as shown in Figure~\ref{fig:breakdown}, reveals critical scalability limitations in existing CPU/GPU implementations.
When datasets scale from 1M to 1B entries, we identify that: 1) IVFPQ’s runtime bottleneck shifts with dataset scale. For example, when dataset scales from 1M to 1B, the performance bottleneck of CPU implementation shifts from the LUT construction stage, which is mainly \emph{compute-intensive}, to the distance calculation stage, which is mainly \emph{memory-bound}. Thus, existing performance optimizations for million-scale ANNS~\cite{vectorsearch-sc23, abdelhadi2019accelerated} may not be suitable to billion-scale. 
2) At billion-scale, both CPU and GPU implementations face various limitations due to architectural mismatch. CPUs become memory bandwidth-limited, struggling to fetch compressed vectors from DRAM for distance calculation. On the other hand, although GPUs excel at parallel distance calculations, they stall during the low-parallelism top-k stage (64\% of runtime), causing resource waste.

\textbf{Opportunities of PIM.}
To bridge this gap, we adopt a new hardware architecture, namely the UPMEM Processing-in-Memory (PIM) hardware~\cite{upmemstat}, which embed computation within memory to minimize data movement and enhance efficiency~\cite{devaux2019true}. Unlike existing PIM products~\cite{asgari2021fafnir, lee20223d, yuan2025fanns, haikun-atc24, cxl-anns},
UPMEM PIM is a standard DIMM device that features multi-threaded DPU cores with direct access to memory, significantly improving performance for data-intensive applications. While recent studies have explored the use of PIM/near-memory technology for ANNS, they are either designed for million-scale~\cite{xu2023proxima} or rely on simulated architectures~\cite{lee2022anna,liang2024hyqa}. \emph{To the best of our knowledge, this is the first study to leverage practical PIM hardware to enhance billion-scale ANNS performance. }

\textbf{Challenges.}
While seem promising, leveraging UPMEM PIM for billion-scale ANNS introduces three challenges: \circled{1} UPMEM’s high aggregated bandwidth depends on evenly distributed workloads across its PIM chips. However, real-world datasets exhibit severe imbalances, skewing compute and memory demands across PIM chips. \circled{2} UPMEM's unique hardware architecture, e.g., distributed DPU cores, tiered memory, and lack of inter-DPU communication, requires rethinking IVFPQ's pipeline to maximize efficiency. 
\circled{3} Despite reducing data movement, UPMEM’s DPUs (350 MHz RISC cores) lack the raw compute power of CPUs/GPUs. This necessitates the design of pruning methods to bypass non-critical computations (e.g., low-impact distance calculations) to maintain low latency while preserving ANN query accuracy.

\textbf{Innovations.}
We propose \emph{UpANNS}, the first framework to enable efficient billion-scale ANNS on commercial PIM hardware.
UpANNS integrates four key innovations to address the aforementioned challenges: \circled{1} We introduce architecture-aware data placement and query scheduling to dynamically balance workloads across PIM chips to maximize  memory bandwidth utilization (Section~\ref{sec::scheduled-model}). \circled{2} We design an efficient thread scheduling approach to fully utilize the multi-threaded computing power on PIM, together with efficient memory management to maximize cache reuse on PIM to further improve the performance (Section~\ref{sec::multi-threading}).
\circled{3} On observing that real-world datasets often contain correlated items in data point vectors, we redesign the encoding in IVFPQ and pre-store frequently accessed item combinations to minimize redundant memory accesses (Section~\ref{sec::encoded-structure}).
\circled{4} We optimize the low-parallelism top-k stage with thread-local top-k heaps and early termination pruning to accelerate the stage on PIM's limited cores (Section~\ref{sec::optimization-strategies}).

We evaluated the performance of UpANNS using seven UPMEM PIM modules and three real-world billion-scale datasets. Results demonstrates three key advantages over conventional architectures: 1) \textbf{Performance}: UpANNS improves the {Query per Second} (QPS) by 4.3x compared to CPU-based Faiss~\cite{faiss}, and is comparable to GPU-based Faiss. 2) \textbf{Energy efficiency}: UpANNS provides 2.3x higher QPS/Watt than GPUs, enabling superior throughput under identical power budgets. 3) \textbf{Scalability}: UpANNS demonstrates near-linear scaling with dataset size, ensuring practical viability for applications like large model serving~\cite{liang2024hyqa}.

\section{Background and Motivation}\label{sec:background}

\subsection{Approximate Nearest Neighbor Search}\label{sec:anns}

\begin{figure}[t]
    \centering
    \includegraphics[width=1\linewidth]{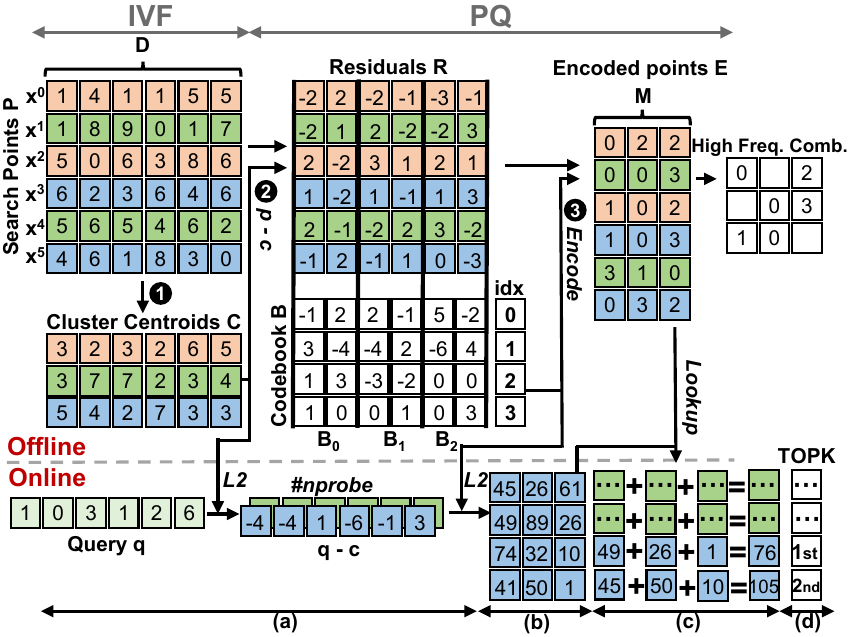}
    \caption{An example of the IVFPQ algorithm (L2 distance).}
    \label{fig:ivfpq}
\end{figure}

Approximate Nearest Neighbor Search (ANNS) algorithms are commonly employed to identify the top-k vectors most similar to a \emph{query vector} within large datasets. Among the various ANNS algorithms, we focus on the Inverted File with Product Quantization (IVFPQ) method~\cite{jegou2010product}, which is one of the most popular ANNS algorithms for its efficiency in billion-scale searches. 
As shown in Figure~\ref{fig:ivfpq}, IVFPQ operates in two phases, namely the \emph{offline preprocessing} phase (top) and the \emph{online querying} phase (bottom). 

{\bf Offline phase} preprocesses dataset to enable rapid searches. First, the Inverted File (IVF) technique 
partitions data points into $|C|$ clusters (e.g., via K-means), and each point is represented as a residual (i.e., the difference between the point and its centroid). Next, Product Quantization (PQ) compresses these residuals by splitting each vector into $M$ subvectors and encoding them into compact codes using a codebook. 
For example, a 128-dimensional vector can be reduced from 512 bytes to 64 bytes using \emph{uint8} encoding with $M=16$, achieving an 8x compression rate. This compressed representation minimizes storage and accelerates distance computations during queries.


{\bf Online phase} processes incoming queries in four stages. (a) \emph{Cluster filtering} computes distances between the query and all cluster centroids, retaining only the $nprobe$ closest clusters to limit the search scope. (b) \emph{Lookup table (LUT) construction} precomputes partial distances between the query and codebook entries for each subvector, eliminating redundant calculations. (c) During \emph{distance calculation}, the algorithm aggregates these precomputed LUT values to approximate the distance between the query and every candidate point in the selected clusters. (d) \emph{Top-K selection} sorts these approximate distances and returns the $k$ nearest neighbors.

\subsection{UPMEM PIM Architecture}\label{sec::arch}



\begin{figure}
    \centering
    \includegraphics[width=0.9\linewidth]{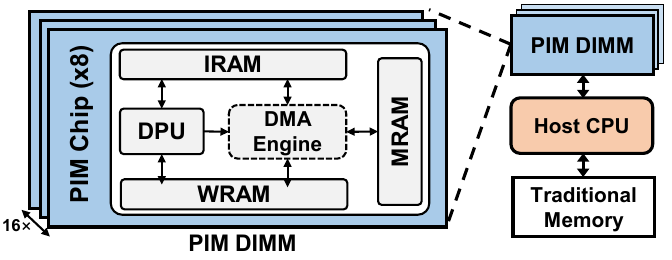}
    \caption{UPMEM PIM architecture.}
    \label{fig:upmem}
\end{figure}


Processing-in-Memory (PIM) represents a promising paradigm for addressing the data movement bottleneck encountered in many large-scale data analytics applications including ANNS. Recently, real PIM hardware has started to enter the market, with UPMEM PIM being the first commercially available solution~\cite{upmemstat,devaux2019true}.
While prior work has demonstrated UPMEM’s utility in recommendation systems~\cite{chen2024updlrm}, graph processing~\cite{cai2024pimpam,giannoula2024}, machine learning~\cite{gogineni2024swiftrl, 10.1145/3656019.3676947} and databases~\cite{10598012, bernhardt2023pimdb}, its potential for ANNS remains unexplored.

A UPMEM module (Figure~\ref{fig:upmem}) is a standard DDR4 DIMM housing 16 PIM chips, each containing 8 DRAM Processing Units (DPUs). Each DPU is a 350 MHz RISC core supporting 24 hardware threads and a 14-stage pipeline to hide memory latency~\cite{upmemusermanual}. The memory hierarchy comprises three tiers, including 1) \emph{MRAM (64MB/DPU)}, which is a slow bulk storage accessed via high-latency DMA transfers; 2) \emph{WRAM (64KB/DPU)}, a fast scratchpad memory with single-cycle access used for active computations and 3) \emph{IRAM (24KB/DPU)} which is the instruction memory for thread execution.

The host CPU orchestrates data transfers to MRAM, which occur concurrently only if buffer sizes across all DPUs are uniform; otherwise, sequential transfers degrade performance. Critically, DPUs lack direct intercommunication and all coordination must route through the host, introducing overhead for collaborative tasks like top-k aggregation. Despite these constraints, UPMEM’s distributed architecture achieves 20 Tb/s aggregate bandwidth~\cite{cai2024pimpam} by parallelizing access across 2,560 DPUs (20 DIMMs), significantly surpassing the memory bandwidth typically available to CPUs.


\subsection{Motivation and Challenges}\label{sec::motivation}
In this subsection, we discuss the opportunities and challenges associated with PIM-based ANNS acceleration.


{\bf Opportunities.}
Scaling ANNS to billion-entry datasets exposes critical inefficiencies in conventional architectures, as highlighted in Figure~\ref{fig:breakdown}.
Specifically, for a dataset with one billion vectors, the distance calculation stage alone requires 250 million random memory accesses per query (with $M=32$, $|C|=4096$ and $nprobe=32$ as in Figure~\ref{fig:breakdown}).
This overwhelms CPU memory bandwidth (e.g., 85 GB/s) and exceeds GPU memory capacity (e.g., 80GB), creating a prohibitive memory wall. UPMEM PIM directly addresses this bottleneck through its 7.2 TB/s aggregated memory bandwidth~\cite{cai2024pimpam}, enabled by parallel access across thousands of DPUs. Unlike GPUs, UPMEM’s standard DIMM form factor supports scalable memory expansion, allowing it to manage datasets far beyond GPU memory limits. By executing computations in-memory, UPMEM eliminates costly data transfers between CPU and memory, making it uniquely suited for memory-intensive ANNS workloads like IVFPQ.


\begin{figure}[t]
    \centering
    \begin{subfigure}[b]{0.155\textwidth}
        \centering
        \includegraphics[width=\textwidth]{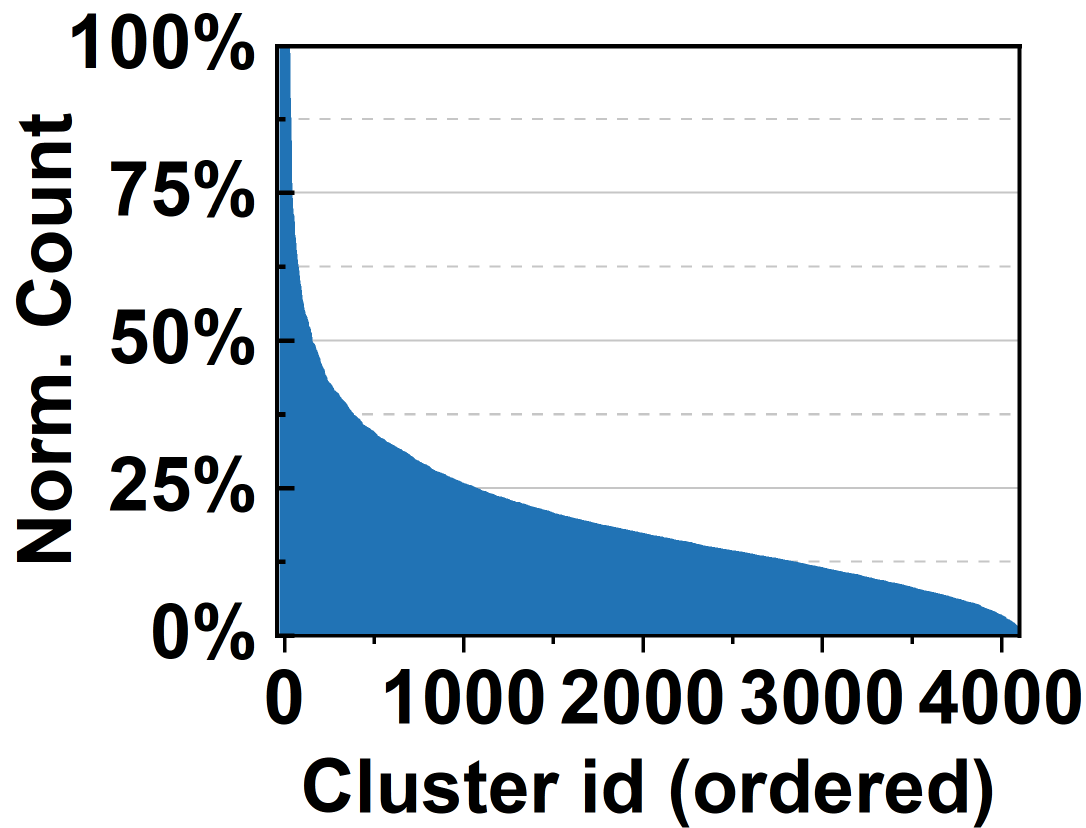}
        \caption{Access frequency}
        \label{fig:motive:a}
    \end{subfigure}
    \begin{subfigure}[b]{0.155\textwidth}
        \centering
        \includegraphics[width=\textwidth]{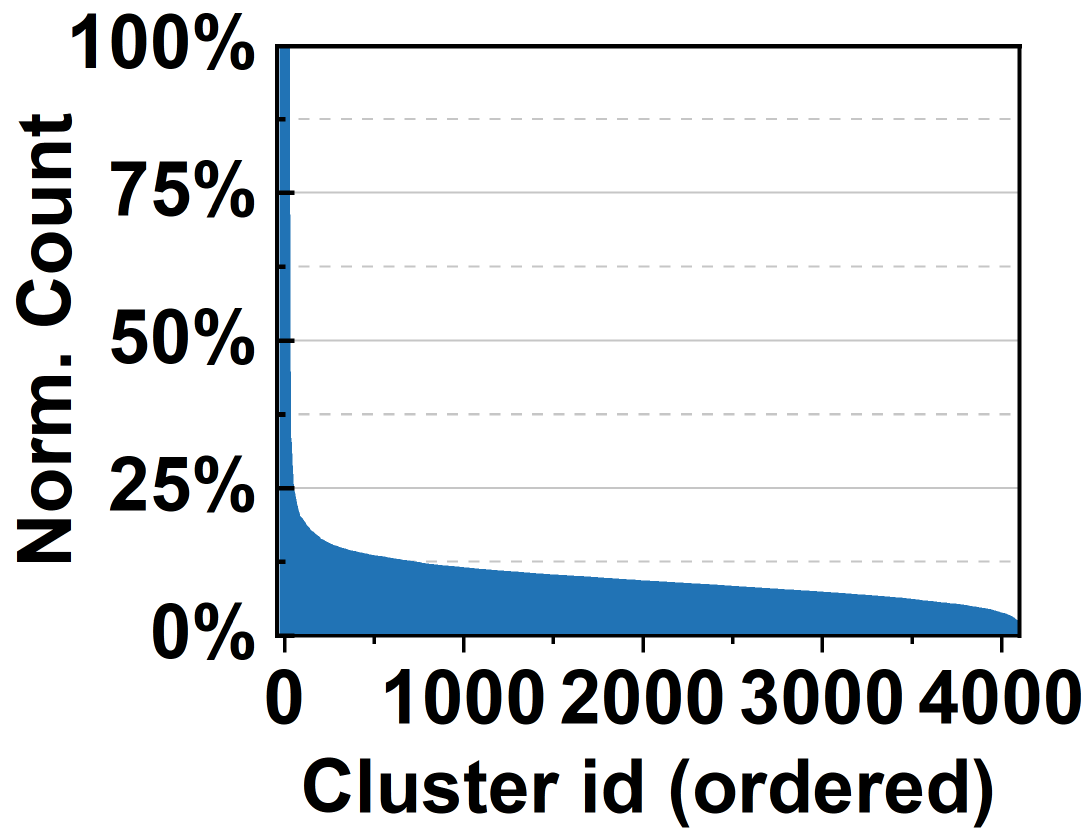}
        \caption{Cluster size}
        \label{fig:motive:b}
    \end{subfigure}
    \begin{subfigure}[b]{0.155\textwidth}
        \centering
        \includegraphics[width=\textwidth]{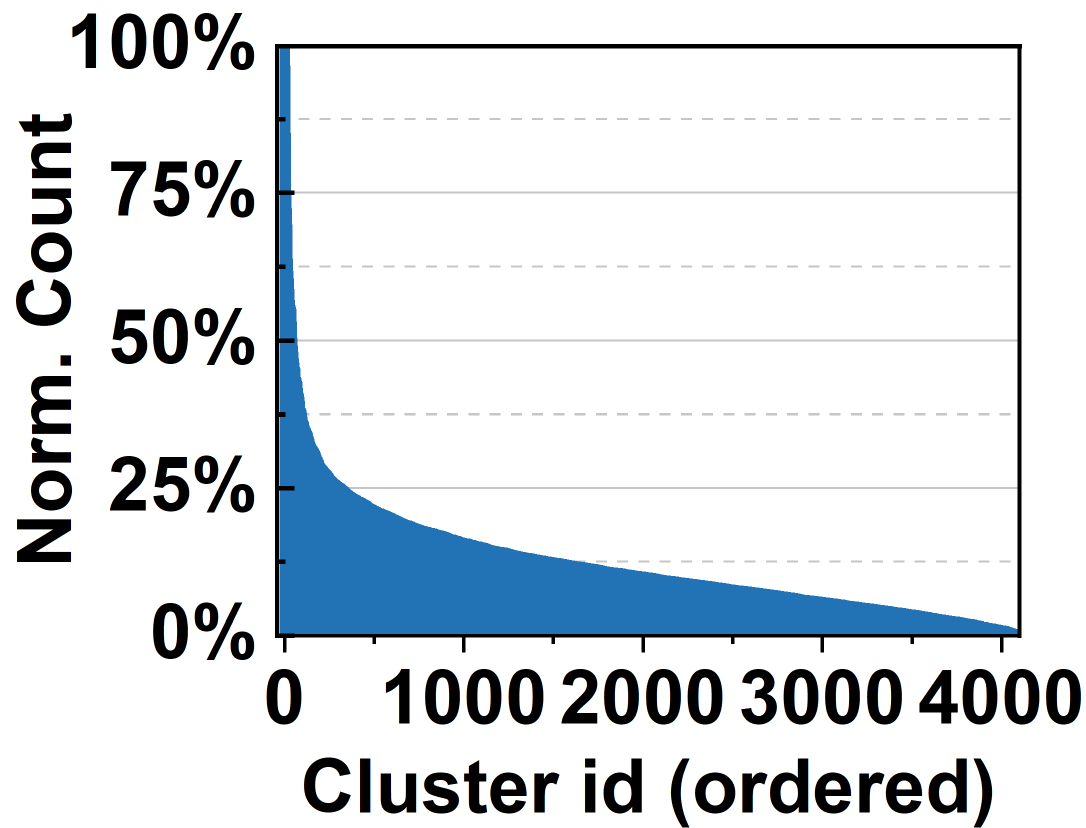}
        \caption{Workload}
        \label{fig:workload}
    \end{subfigure}
    \caption{The distribution of access frequency, cluster size and workload in each cluster in SPACEV1B~\cite{spacev}.}
    \label{fig:powerlaw}
\end{figure}

\begin{figure*}[t]
    \centering
    \includegraphics[width=\linewidth]{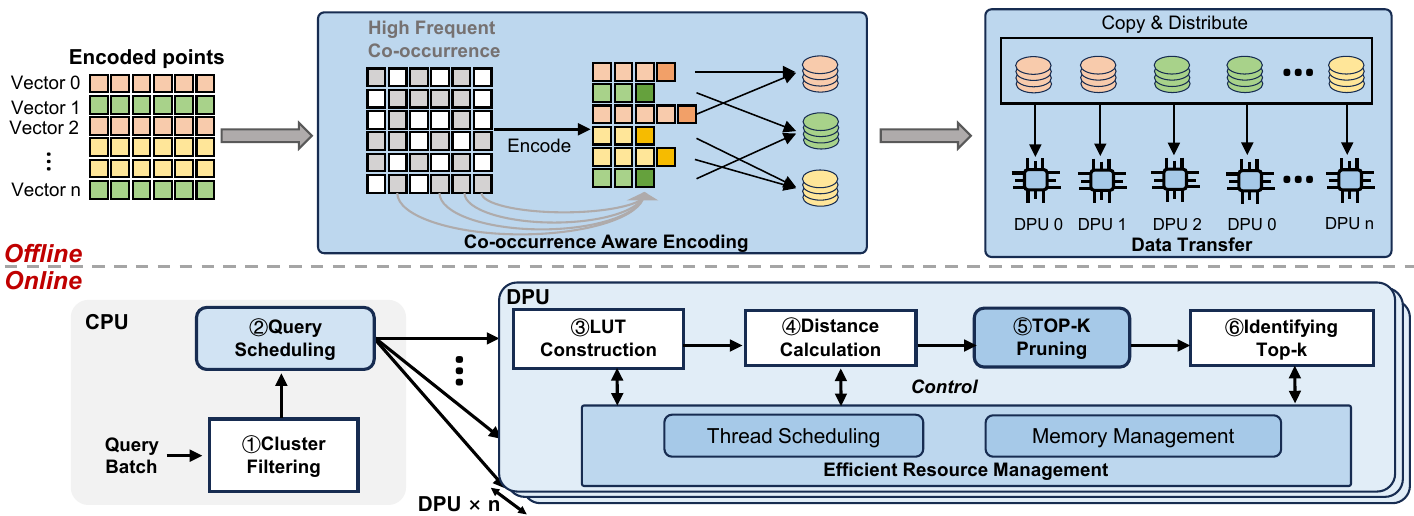}
    \caption{Overview of UpANNS. The grey blocks in each vector represent high frequency co-occurrences in the encoded vectors.}
    \label{fig:system}
\end{figure*}

\textbf{Challenges.}
While UPMEM PIM offers great potential, realizing its benefits for billion-scale ANNS requires overcoming three key challenges tied to its architecture and workload characteristics:


\emph{Challenge 1: Imbalanced data access pattern.}
IVFPQ’s cluster-centric design interacts poorly with UPMEM’s distributed DPUs. Real-world datasets exhibit extreme skew: popular clusters receive 500x more queries than others (Figure~\ref{fig:motive:a}), and large clusters contain $10^6$x more data points than others (Figure~\ref{fig:motive:b}). 
This imbalance creates ``hot'' DPUs overwhelmed by frequent and large cluster accesses while others idle, wasting UPMEM’s parallel bandwidth. 
Compounding this, DPUs lack direct communication. Data communications between DPUs have to go through slow host CPU, negating PIM’s latency benefits.



\emph{Challenge 2: Complicated PIM resource management.}
UPMEM provides parallel structure between DPUs. Each DPU equips with 24 hardware threads and 14-stage pipeline, with exclusive access to its local memory. How to adapt IVFPQ to fully utilize the multi-threads and leverage the 14-stage pipeline to maximally reduce the pipeline bubble and hide the memory read latency is a challenging problem. 
In addition, the fast WRAM on DPU has a limited capacity of 64KB only. Due to the significant bandwidth difference between WRAM and MRAM, we should maximize the amount of data stored in WRAM to improve performance.
However, UPMEM lacks a Memory Management Unit (MMU) to virtualize its physical memory. DPUs use physical addresses when accessing memory~\cite{hyun2024pathfinding}, which prohibits efficient utilization of the limited cache space. 

\emph{Challenge 3: Limited power of UPMEM PIM.}
While UPMEM reduces data movement, its DPUs lack the computational density of CPUs/GPUs. Processing millions of memory accesses per query on 350 MHz cores risks prohibitive latency. This necessitates accuracy-aware pruning strategies to skip non-critical computation without degrading query accuracy.

\section{System Overview}

This paper presents UpANNS, an efficient system that accelerates ANN search using UPMEM PIM hardware.
\sysname{} incorporates four optimizations, targeting specific challenges mentioned above. 

\textbf{Opt1: PIM-Aware Workload Distribution.}
To mitigate skewed cluster access patterns (\emph{challenge 1}), we dynamically replicate and distribute IVF clusters across DPUs during offline preprocessing. High-demand clusters are replicated to avoid hotspots, while spatially proximate clusters are co-located on the same DPU to minimize costly inter-DPU communication. This ensures balanced memory accesses and maximizes the utilization of UPMEM’s 7.2 TB/s aggregated bandwidth.

\textbf{Opt2: PIM Resource Management.}
With UPMEM's unique hardware architecture (\emph{challenge 2}), we propose two optimizations to fully utilize its limited resources: 1) Thread scheduling: we utilize 24 hardware threads per DPU to overlap MRAM transfers with computations via a 14-stage pipeline; 2) Memory management: we carefully reuse WRAM buffers across IVFPQ stages (e.g., codebook, LUT to encoded vectors) to minimize MRAM accesses (Figure~\ref{fig:multithread}).

\textbf{Opt3: Co-occurrence Aware Encoding.}
Targeting \emph{challenge 3}, we identify frequent element combinations in encoded vectors and precompute partial sums during offline processing. This can greatly reduce online memory accesses without compromising accuracy.

\textbf{Opt4: Top-K Pruning.}
We further address \emph{challenge 3} by eliminating redundant computations. For the bottlenecked top-k selection stage, we combine thread-local heaps with early termination, which enables skipping 68\% of redundant comparisons and accelerates the final stage by 3.1x.

Figure~\ref{fig:system} shows the optimized workflow of IVFPQ. During offline phase, clusters are encoded (Opt3), replicated and distributed (Opt1). During online phase, queries are scheduled to balanced DPUs (Opt1), and processed using WRAM-optimized pipelines (Opt2) and pruned top-k selection (Opt4).

\section{Technical Details}


\subsection{PIM-Aware Workload Distribution}\label{sec::scheduled-model}
To address the severe workload imbalance caused by skewed cluster access patterns, UpANNS integrates two synergistic strategies: \emph{offline data placement} and \emph{online query scheduling}. 
The data placement task selectively replicates and distributes clusters across DPUs based on their access frequency patterns and sizes, enhancing system flexibility for diverse query workloads. Meanwhile, the query scheduling task efficiently maps filtered clusters from batch queries to appropriate DPUs, dynamically balancing workloads at runtime.

\subsubsection{\textbf{Data Placement.}}\label{sec:placement}

The offline data placement strategy optimizes cluster distribution across DPUs based on three three key insights: (1) cluster-level locality preservation reduces partial result transfers, (2) workload skew necessitates adaptive replication, and (3) spatial proximity between frequently co-accessed clusters enables communication optimizations. 
Our data placement strategy carefully considered these three insights to achieve optimal performance and the key steps are shown in Algorithm~\ref{alg:placement}.

Firstly, entire clusters are placed on a single DPU to avoid partial result transfers on the bottlenecked communication between DPUs and the host CPU.
For a cluster $i$, its workload is mainly contributed by the memory-intensive distance calculation stage, which performs excessive memory search in LUTs to calculate distances between a query $q$ and all search points in the cluster. Thus, we can estimate the workload as $W_i=s_i*f_i$, where $s_i$ is the cluster size and $f_i$ is its historical access frequency. Given $n$ DPUs and a batch of queries, the balanced memory accesses per DPU that we need to achieve is $\overline{W}=\frac{1}{n}\sum_{i=1}^mW_i$, where $m$ is the number of clusters.

\floatname{algorithm}{Algorithm} 
\renewcommand{\algorithmicrequire}{\textbf{Input:}}  
\renewcommand{\algorithmicensure}{\textbf{Output:}}  
\begin{algorithm}[t]\scriptsize
\caption{Data Placement for Cluster $i$}\label{alg:placement}
\begin{algorithmic}[1]
\Require {$ndpu$: the number of available DPUs; $s_i$: \#vectors in cluster $i$; \newline $f_i$: the access frequency of cluster $i$; $\overline{W}$: the average workload per DPU; \newline $MAX\_DPU\_SIZE$: the maximum number of vectors each DPU can have;}
\Ensure {$dpu\_id$};
\State $d\_id \gets ndpu$, $thld \gets 1.0$, $count \gets 0$ and $rate \gets 0.02$;
\State $ncpy = \lceil {s_i*f_i}/{\overline{W}}\rceil$; \Comment{$ncpy$ denotes \#DPUs that cluster $i$ is distributed onto;}
\State $w_i = {s_i*f_i}/{ncpy}$; \Comment{$w_i$ denotes the per DPU workload of cluster $i$;}

\While{$ncpy > 0$}    
    \If{$W[d\_id] + w_i \leq \overline{W} * thld$ \textbf{and} $S[d\_id] + s_i \leq MAX\_DPU\_SIZE$} 
        \State $count \gets 0$;
        \State $dpu\_id.\text{push}(d\_id)$, $ncpy \gets ncpy - 1$;
    \Else
        \State $count \gets count + 1$;
    \EndIf
    \State $d\_id \gets (d\_id + 1) \mod ndpu$;
    \If{$count == ndpu$} \Comment{no suitable DPU found;}
        \State $thld \gets thld + rate$; \Comment{enlarge threshold to loosen the workload balance constraint;}
    \EndIf
\EndWhile
\end{algorithmic}
\end{algorithm}

Secondly, cluster access frequency is highly skew, as shown in Figure~\ref{fig:powerlaw}. Thus, some clusters may have workloads significantly exceeding $\overline{W}$. We create $ncpy$ copies for these high-demand clusters (Line 2) and distribute them across multiple DPUs. 
During placement, replicas are assigned to DPUs with the least residual capacity, iteratively increasing the acceptable workload threshold $thld$ until all replicas are placed (Lines 5–12).
At the end of this algorithm, each cluster is placed on a list of $ncpy$ DPUs.

Thirdly, clusters selected for a single query are typically in close proximity (e.g., neighboring centroids). We co-locate such clusters on the same DPU, which allows local aggregation of partial top-k results during multi-cluster queries, dramatically cutting inter-DPU communication overhead.

\floatname{algorithm}{Algorithm} 
\renewcommand{\algorithmicrequire}{\textbf{Input:}}  
\renewcommand{\algorithmicensure}{\textbf{Output:}}  
\begin{algorithm}[t]\scriptsize
\caption{Query Scheduling for a batch $Q$}\label{alg:scheduling}
\begin{algorithmic}[1]
\Require {$s_i$: \#vectors in cluster $i$; $C$: centroid vectors;\newline $M$: the cluster to DPU mapping generated by Algorithm~\ref{alg:placement};}
\Ensure {$Assigned$: the query scheduling result;}
\State $W[i] \gets 0$ \textbf{for} $i \in \{0, \ldots, ndpu-1\}$; \Comment{$W[i]$ denotes the workload on DPU $i$}
\For{$i \in [0, |Q|-1]$}
    \State $F[i] \gets \text{cluster\_filtering}(Q[i])$; \Comment{IDs of selected clusters for query $i$}
    \For{each cluster $j$ in $F[i]$ \textbf{and} $M[j].size = 1$ } \Comment{clusters with one replica}
        \State $Assigned[M[j][0]].\text{push}(\langle i, j\rangle)$;\Comment{Schedule cluster $j$ of query $i$ to DPU $M[j][0]$}
        \State $W[M[j][0]] \gets W[M[j][0]] + s_j$;
        \State Remove cluster $j$ from $F[i]$;
    \EndFor
\EndFor
\State Sort all clusters in $F$ in descending order according to their sizes;
\For{each cluster $c$ in $F$}
    \State $dpus \gets M[c]$; \Comment{the list of DPUs containing replicas of cluster $c$}
    \For{each query $Q[i]$ \textbf{and} $F[i]$ contains $c$}
        \State \textbf{find} $j$ \textbf{with} $\min(W[dpus[j]] + s_{c})$; \Comment{find the least loaded DPU}
        \State $Assigned[dpus[j]].\text{push}(\langle i, c\rangle)$;
        \State $W[dpus[j]] \gets W[dpus[j]] + s_c$;
        \EndFor
\EndFor
\end{algorithmic}
\end{algorithm}

\begin{figure*}[t]
\begin{minipage}{0.65\textwidth}
\centering
    \includegraphics[width=\linewidth]{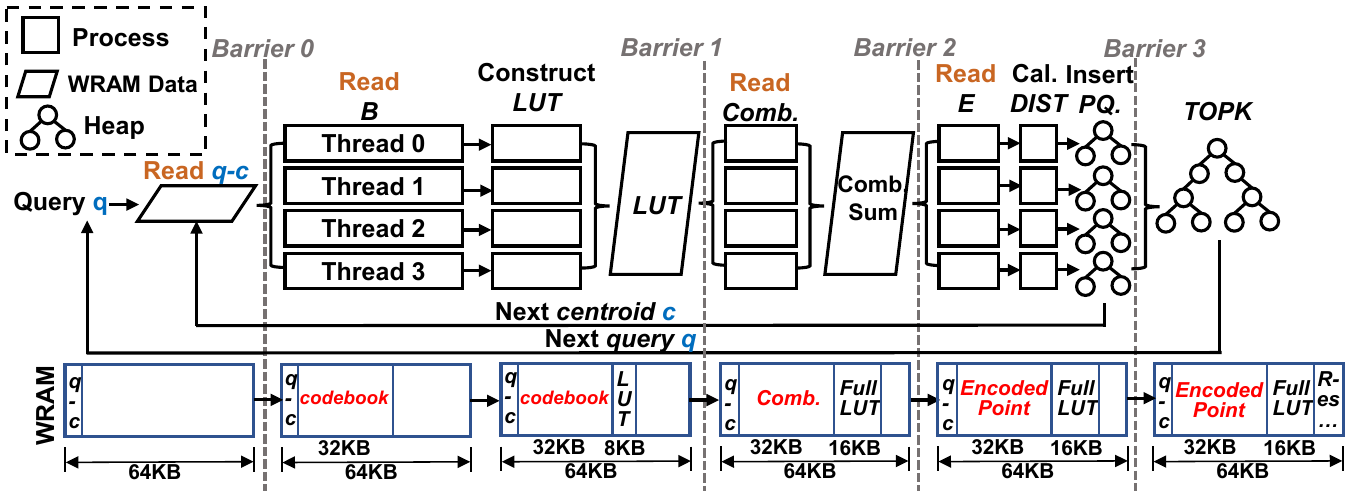}
    \caption{Parallel processing of each cluster inside a single DPU. Red texts represent that the WRAM space is reused.}
    \label{fig:multithread}
\end{minipage}
\hfil
\begin{minipage}{0.25\textwidth}
\centering
    \includegraphics[width=\linewidth]{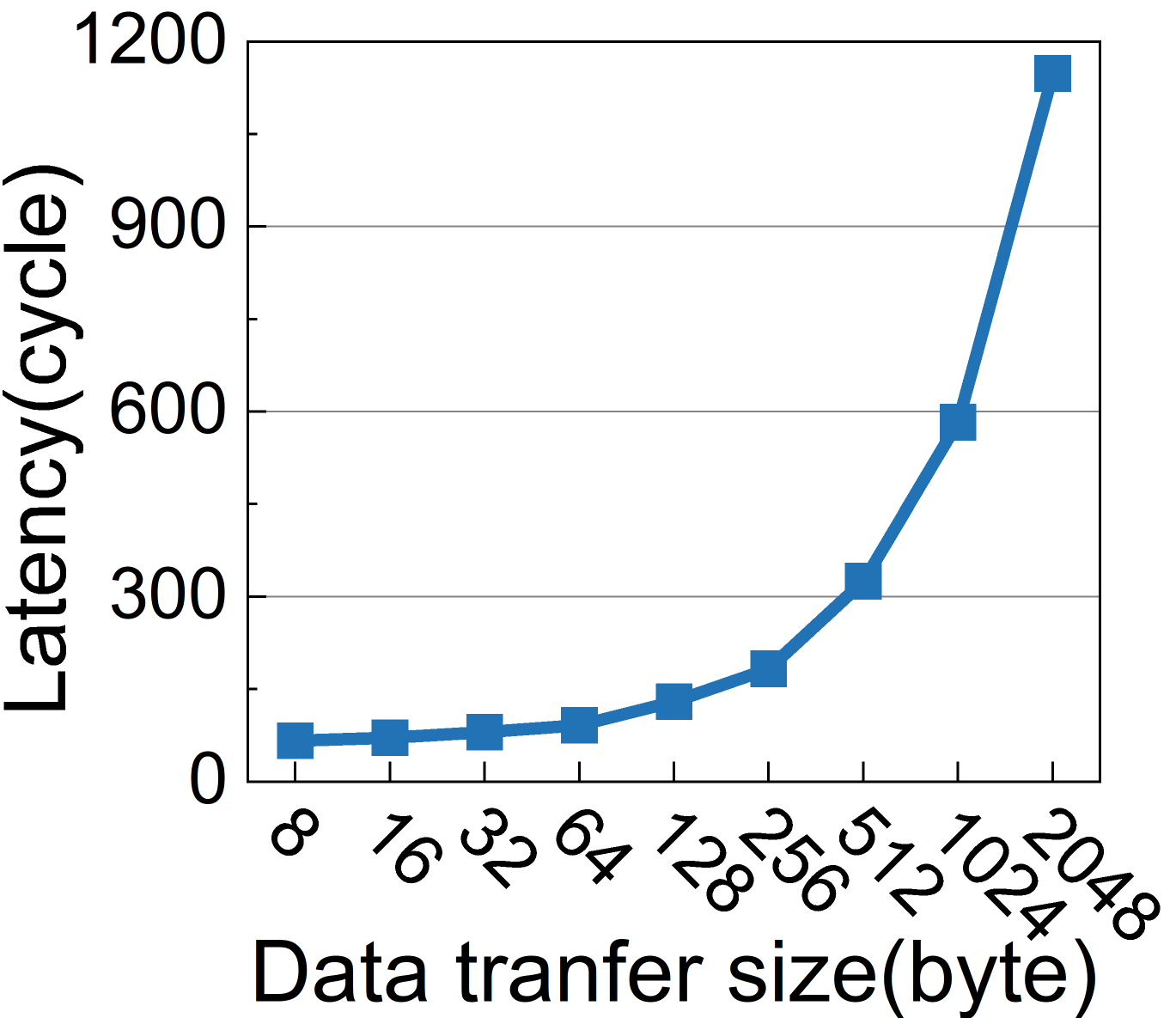}
    \caption{MRAM read latency}
    \label{fig:mram:latency}
\end{minipage}
\end{figure*}

\subsubsection{\textbf{Query Scheduling.}}
At runtime, the host CPU assigns queries to DPUs using a greedy scheduling algorithm (Algorithm~\ref{alg:scheduling}) to achieve workload balance dynamically.
Each query selects $nprobe$ clusters by comparing the centroid vectors with the query vector, thus is relatively light-weighted compared to the other online stages. We execute cluster filtering and query scheduling on the host CPU and send the scheduling results to PIM for query searching.

Since some popular clusters are replicated and distributed across DPUs, the query scheduling algorithm dynamically selects the appropriate cluster replicas to ensure workload balance. 
Specifically, we first schedule the clusters with only one replica (Line 4-5). Next, we update the workload on each DPU (Line 6). Clusters with multiple replicas are ordered by size (descending) and assigned to the least-loaded DPU (Line 8-14). This ensures larger clusters do not overload individual DPUs.
Algorithm~\ref{alg:scheduling} runs at runtime with a complexity of $O(|Q| \times nprobe)$, where $Q$ is the set of queries. Since both $|Q|$ and $nprobe$ are significantly smaller than billion-scale datasets, the overhead of this algorithm is negligible. 

Due to hardware limitations preventing direct DPU-DPU communication, PIM faces challenges in handling changing query patterns. UpANNS targets applications such as retrieval-augmented LMs and recommendation systems, where query patterns typically change regularly (e.g., every few days~\cite{vectorsearch-sc23}) and incrementally. To handle such changes, UpANNS implements an \emph{adaptive} approach: (1) adjusting the number of cluster copies (Section~\ref{sec:placement}) to accommodate minor query pattern changes in the short term, and (2) performing full data relocation to address major pattern shifts over longer time periods. Furthermore, our experiments demonstrate that the PIM hardware achieves significantly higher query throughput than CPU implementations even with random data distribution, highlighting the substantial advantages of PIM architectures.

\subsection{PIM Resource Management}\label{sec::multi-threading}

To maximize UPMEM’s limited per-DPU resources (small WRAM cache and 350 MHz DPU cores), UpANNS employs two synergistic optimizations: \emph{thread scheduling} and \emph{memory management}. These ensure full utilization of PIM’s parallel compute and memory bandwidth while mitigating latency bottlenecks.

\subsubsection{\textbf{Thread Scheduling.}}
The UPMEM architecture provides a parallel structure between DPUs. Further, each DPU’s 24 hardware threads and 14-stage pipeline enable fine-grained parallelism, but WRAM constraints (64KB) restrict concurrent cluster processing. 

To efficiently utilize the multi-threads, we can explore hardware parallelism at various levels: inter-queries, between different clusters of a single query, or intra-cluster.
Note that multiple threads operating on the same DPU share the limited 64KB WRAM. Since accessing data from WRAM is much faster than from MRAM, we try to fit all data needed during thread execution in WRAM, including the codebook, LUT and encoded data in selected clusters. 
The codebook size can be estimated as $D \times 256$, where $D$ represents the dimension of data points, amounting to 32KB for the SIFT dataset.
The LUT size can be estimated as $M \times 256 \times sizeof(uint16)$, which is 8KB when the dimension of encoded points $M$ is 16. 
If we pursue parallelism at the query or cluster level, the WRAM space will be over-utilized since the combined size of the LUTs for more than four clusters will easily exceed 64KB. Given this hardware constraint, \emph{we implement intra-cluster parallelism}, where threads collaborate on a single cluster at a time.


On each DPU, multiple threads execute concurrently to accelerate the processing of a single cluster. Figure~\ref{fig:multithread} illustrates our multi-threading strategy.
Specifically, in the LUT construction stage, threads concurrently fetch codebook segments \emph{from MRAM into WRAM}, constructing partial lookup tables (LUTs). For $M=16$ subvectors, this reduces LUT build time by 4.8x compared to single-threaded execution.
{After the LUT is constructed, the partial sum of high-frequency co-occurrence combinations are computed using the LUT (details introduced in Section~\ref{sec::encoded-structure}). The partial sums and the LUT will be combined to construct the full LUT.}
Once the full LUT is constructed, multiple threads concurrently read the encoded points $E$ \emph{from MRAM to WRAM} to calculate the partial distances between the query and search points. 
Pipelining is enabled to overlap MRAM reads and distance calculations to hide the high latency.
Each thread maintains a thread-local priority queue (PQ), implemented as a max heap with a size of $k$, to store the local top-k partial distances. After a thread finishes the partial top-k insertion, it moves on to process the next cluster.

Four \emph{barriers} are introduced to synchronize the threads and guarantee the correctness of processing. \emph{Barrier 0} prevents premature LUT updates while some threads are still calculating DIST based on LUT. \emph{Barrier 1} ensures the LUT is fully constructed before combination sums are created. \emph{Barrier 2} secures the updated LUT and combination sums, preventing erroneous reads. \emph{Barrier 3} confirms that all threads have completed the distance calculations and result insertions for the current query, allowing for aggregating the final top-k from the thread-local priority queues.

The number of threads (\#threads) is an important parameter in the implementation. A DPU can support up to 24 threads. However, since the size of the MRAM-WRAM transfer must be a multiple of 8, at least equal to 8 and not greater than 2048, each thread requires up to 2KB WRAM space to store the search points read from MRAM. Next, we discuss how to efficiently reuse the limited WRAM capacity to allow more threads for better performance.

\subsubsection{\textbf{Memory Management.}}
Due to the significant bandwidth difference between WRAM and MRAM, we aim to maximize the data stored in WRAM.
The UPMEM DPU lacks a Memory Management Unit (MMU) to virtualize its physical memory, and the DPU is constrained to the limited physical memory capacity. 
Therefore, we propose a WRAM reuse strategy specifically tailored to the query processing stages on PIM. This strategy enables threads to operate in a larger address space than the physical WRAM space available.  

Figure~\ref{fig:multithread} illustrates the reuse strategy using the SIFT dataset as an example. During the LUT construction stage, we use multiple threads to read the codebooks (32KB) and compute the corresponding entries in the LUT (8KB). Once the complete LUT is obtained, we use the same capacity to construct the sum of the combinations (8KB). Since the codebooks are not used in the following stages, they can be safely overwritten to conserve space. As a result, the total space required for the codebooks and the full LUT is only 48KB. 
Once the full LUT is obtained, we proceed to the costly distance calculation stage, which needs to fetch encoded point from MRAM to WRAM. To optimize this process, we reuse the WRAM space occupied by codebooks to allow more threads to load encoded data points concurrently. In the example shown in Figure~\ref{fig:multithread}, we utilize 16 threads consuming 32KB of memory. 

The read latency of MRAM does not increase linearly with size, as shown in Figure~\ref{fig:mram:latency}. The latency increases slowly as data size grows from 8B to 256B and increases almost linearly beyond 256B. This suggests that smaller MRAM read sizes (under 256B) yield greater benefits. Larger MRAM reads can consume significant WRAM space with minimal returns. We explore an optimal MRAM read size (buffer size) to enhance overall efficiency. 

\subsection{Co-occurrence Aware Encoding}\label{sec::encoded-structure}

The limited computational and memory resources of individual DPUs necessitate aggressive workload reduction without sacrificing accuracy. 
Existing works have proposed various workload pruning techniques using distance bounds~\cite{andre2016cache,gao2023high} and machine learning~\cite{haikun-atc24}. These methods are complementary to UpANNS.

Encoded points are indices pointing to codebooks, resulting in a limited value range (typically within [0, 255]). This property allows  frequent co-occurring element combinations. For instance, in the SIFT1B dataset, the triplet (1, 15, 26) appears in 5.7\% of vectors. 
During distance calculation stage (Figure~\ref{fig:ivfpq}(c)), each element in an encoded vector triggers a LUT lookup, summing to compute the final distance.
By caching the partial sums of the high-frequency combinations, we can reduce redundant memory accesses and computations during distance calculation at runtime, hence improving the querying performance.

\begin{figure}[t]
    \centering
    \includegraphics[width=1\linewidth]{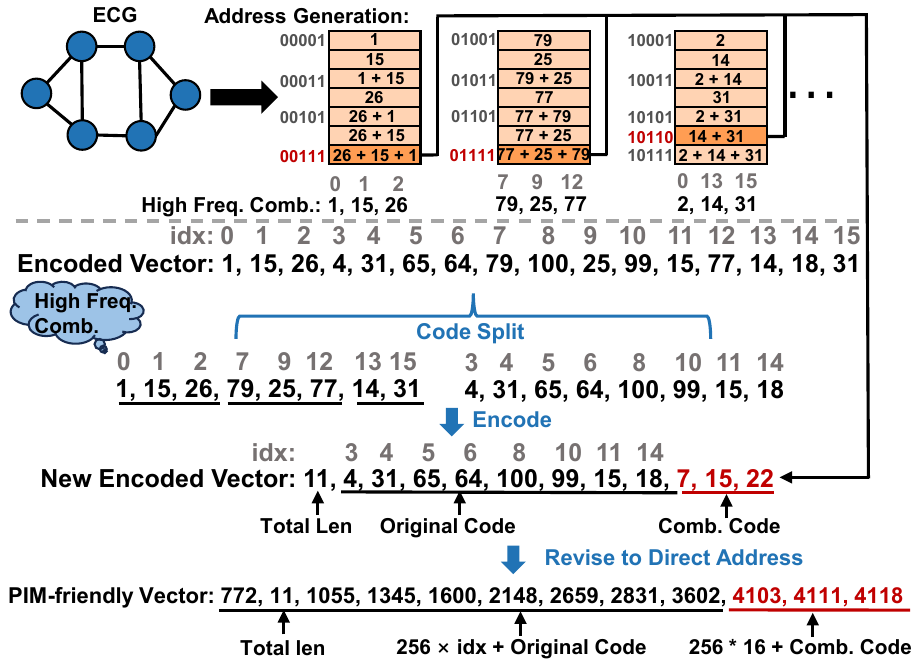}
    \caption{An example of co-occurrence aware encoding.}
    \label{fig:new-code}
\end{figure}

Leveraging the above observations, we propose a \emph{co-occurrence aware data encoding} strategy. Specifically, we first identify high‐frequency co‐occurring element combinations in encoded points using an Element Co‐occurrence Graph (ECG). The ECG is constructed such that nodes represent elements and edges represent co‐occurrence relationships, with edge weights corresponding to co‐occurrence frequencies. For each cluster, we select the top-$m$ most frequent combinations of length 3. Longer combinations can be selected if a larger cache size (e.g., WRAM capacity) is available. The parameter $m$ is determined by the WRAM size; by default, we set $m$=256. 
Our goal is to cache the partial sums of these combinations to avoid redundant computations during online searches. However, since distances are unknown during the offline phase, partial sums cannot be precomputed. To address this, we reserve buffer space in WRAM to store partial sums generated after the LUT construction stage. We preallocate this buffer space and assign memory addresses to each partial sum. These addresses are then used to re‐encode the IVFPQ‐encoded points. Figure~\ref{fig:new-code} illustrates this process with a detailed example.

\textbf{First}, as shown on the top of the figure, we identify three co-occurrence element sets, namely (1, 15, 26), (79, 25, 77) and (2, 14, 31).
Note that, we must also take into account the positions where these occurrences happen. 
For example, when utilizing the cached partial sum of the combination (1, 15, 26), it is crucial that these elements appear in the columns (0, 1, 2) respectively to make the sum meaningful.
\textbf{Second}, we calculate the cache address for the partial sum of all possible combinations in the co-occurrence sets. 
\textbf{Third}, we use the generated cache address to re-encode an encoded vector, which has 16 dimensions and each dimension is a $uint8$ value in the 0-255 range.
This vector contains the high-frequency combinations (1, 15, 26), (79, 25, 77) and (14, 31) on the correct positions. According to the pre-computed cache address, we can find the partial sum of the three combinations at 0x00111 (7), 0x01111 (15) and 0x10110 (22) respectively.
Thus, in the new encoded vector, we use the address of the partial sum to replace the original elements in the vector. Specifically, the new encoded vector contains three parts. The first digit represents the total length of the vector, followed by the original codes which are not identified as high-frequency co-occurred elements and the combination codes which record the cache address of the partial sum. 

During online phase, we calculate the partial sums using constructed LUT and store the sums in WRAM according to pre-arranged layout. For example, for combination (1, 15, 26) on position (0, 1, 2), the partial sum is calculated as $LUT[1+0\times256] + LUT[15+1\times256] + LUT[26+2\times256]$, and stored at address 0x00111.
By reusing the cached partial sum, we can greatly reduce the computation and memory accesses during distance calculation stage.

Note that, during partial sum calculation, we need multiplication operations to access the LUT (e.g., $15+1\times256$). However, existing work~\cite{gomez2021benchmarking} shows that multiplication operations on PIM are less efficient. 
To mitigate this issue, as shown in Figure~\ref{fig:new-code}, we further modify the new encoded vector by converting the original code and combination code into \emph{direct addresses}. 
For the original code, we calculate the direct address of the LUT by $256 \times idx + original\_code$. For the combination code, we directly use the cache address with an offset of the LUT size ($256 \times 16$). The length of the new encoded vector is stored in the second digit, which is 11 in this example. The reason for storing in the second digit is to ensure that the length of the new encoded vector is no more than the original length. If the new vector doesn't have a high frequency combination, the length doesn't need to be stored and the second digit is larger than $256 \times 1$. Otherwise, the length is stored in the second digit and the second digit is less than $256 \times 1$. We could identify the second digit to determine the length of the encoded vector.

In the example, our co-occurrence aware encoding reduces the length of the encoded vector from 16 to 12, achieving a 25\% reduction in length. 
Our empirical studies indicate that, a higher length reduction rate corresponds to a greater decrease in distance calculation time, as demonstrated in Figure~\ref{fig:cache-exp}.

\subsection{Top-K Pruning}\label{sec::optimization-strategies}
The above optimizations try to alleviate the memory bottleneck in the LUT construction and distance calculation stages.
In this subsection, we look at the performance issue in the top-k selection stage to further enhance UpANNS efficiency.

\begin{figure}
    \centering
    \includegraphics[width=\linewidth]{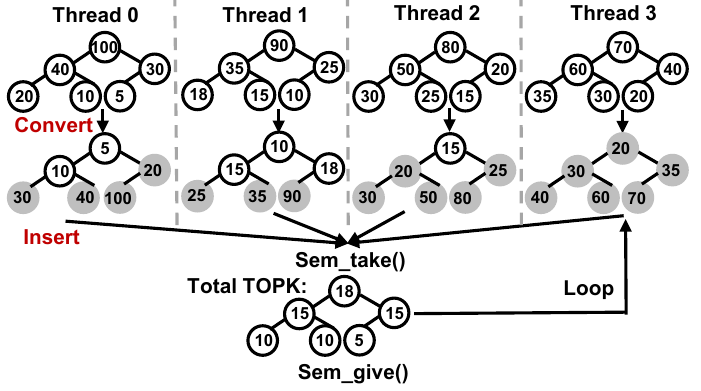}
    \caption{Top-K pruning in DPU. Grey nodes are pruned.}
    \label{fig:heap}
\end{figure}

Recall from Figure~\ref{fig:multithread}, each thread maintains a thread-local priority queue for local top-k. We get the total top-k on the current DPU by aggregating the local top-k results from different threads.
Directly transferring all local top-k to the CPU would lead to too much communication between the CPU and DPUs. 
Instead, we propose to efficiently insert the values from the local top-k queues into the total top-k queue on the current DPU, with pruning enabled.
Figure~\ref{fig:heap} shows a detailed example.


Considered four threads, each maintaining a priority queue of six elements. A max heap is employed to construct the priority queues. Once all four threads have completed their maintenance of the max heaps (Barrier 3), they can use semaphores to concurrently insert values from their thread‐local heaps into the total top‐k. 
To minimize unnecessary insertion, we propose to convert the thread-local max heaps into min heaps. We use the \emph{sem\_take()} and \emph{sem\_give()} semaphore functions to manage concurrent insertions of the top element from each min heap into the total result max heap.
For any thread‐local min heap, if the top value is greater than the maximum value in the total max heap, it indicates that the remaining values in the thread‐local heap cannot contribute to the overall top‐k and can therefore be pruned.

\section{Evaluation}\label{sec::exp}
\begin{table*}[t]\small
\resizebox{1\textwidth}{!}{%
\begin{tabular}{c|c|c|c|c|c}
\hline
\textbf{Harware} & \textbf{Specifications}                                                                  & \textbf{Approx. Price} & \textbf{Memory capacity} & \textbf{Peak Power} & \textbf{Bandwidth}  \\ \hline
\textbf{CPU~\cite{cpu}}     & \begin{tabular}[c]{@{}c@{}}2xIntel Xeon Silver 4110@2.10GHz, 4xDDR4 DRAM\end{tabular} & 1,400USD           & 128 GB                   & 190W           & 85.3 GB/s                    \\ \hline
\textbf{GPU~\cite{gpu}}     & Nvidia A100 PCI-e 80GB                                                                   & 20,000USD          & 80 GB                    & 300W           & 1935 GB/s                   \\ \hline
\textbf{PIM~\cite{falevoz2023energy}}     & 7xUPMEM PIM (896 DPUs)                                                                    & 2,800USD           & 56 GB                    & 162W           & 612.5 GB/s                  \\ \hline
\end{tabular}%
}
\caption{Specifics of evaluated hardware architectures.}
\label{tab:setup}
\end{table*}
\subsection{Experimental Setup}

\textbf{Compared Baselines.} 
We compare UpANNS with three baselines: PIM-naive, Faiss-CPU and Faiss-GPU. PIM-naive is the naive implementation of IVFPQ on PIM with our PIM resource management strategy. Faiss-CPU and Faiss-GPU adopt the CPU- and GPU-based implementations of IVFPQ from Faiss library~\cite{faiss}, the most popular ANNS library developed by Meta. 

While there are other advanced works to accelerate ANNS methods, their methodologies or hardware compatibility differ fundamentally from UpANNS. For example, FANNS~\cite{vectorsearch-sc23} accelerates IVFPQ using FPGAs but faces device memory constraints, limiting scalability to billion-scale datasets. Some studies~\cite{lee2022anna} optimizes IVFPQ via simulator-based architectures, thus are incomparable to UpANNS, which targets real-world hardware. To the best of our knowledge, no accessible PIM-based ANNS implementation exists for direct comparison. For example, Proxima~\cite{xu2023proxima} estimated its performance using an in-house simulator. To bridge this gap, we implemented a rigorous naive PIM baseline (PIM-naive) that serves as a foundation for UpANNS.
Notably, the above works~\cite{lee2022anna,vectorsearch-sc23,xu2023proxima} all use Faiss as the state-of-the-art method, making it a reasonable choice for comparison.



\begin{figure*}[t]
    \centering
    \begin{minipage}{.49\textwidth}
    \hspace*{-1em}
        \includegraphics[width=1\linewidth]{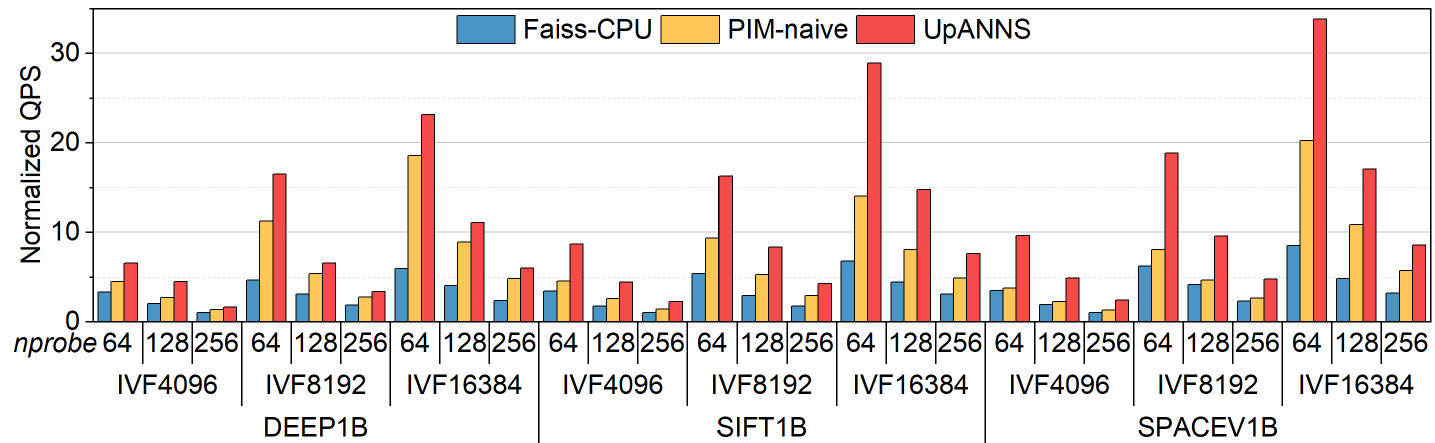}
    \caption{QPS of different ANNS solutions normalized to that of Faiss-CPU when IVF is 4096 and \emph{nprobe} is 256. \emph{nprobe} varies from 64, 128 to 256. \#clusters varies from 4096, 8192 to 16384.}
    \label{fig:overviewpcpu}
    \end{minipage}
    \hfill
    \begin{minipage}{.49\textwidth}
        \includegraphics[width=1\linewidth]{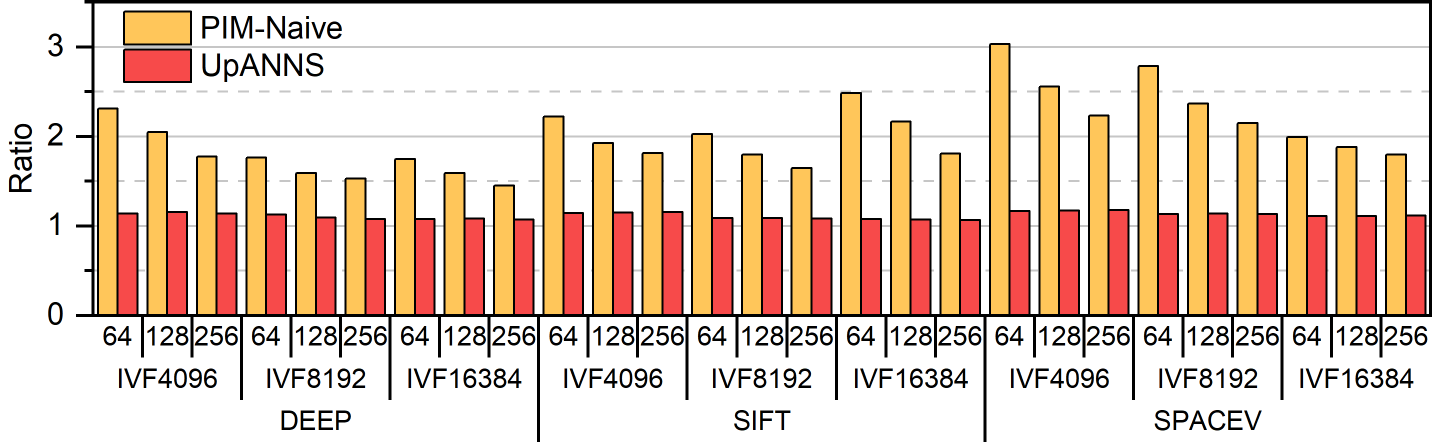}
    \caption{The ratio of maximum process and average process improvement due to the PIM-Aware Workload Distribution strategy under different \emph{nprobe} and IVF settings.}
    \label{fig:scheduled-try}
    \end{minipage}
\end{figure*}

\begin{figure*}[htbp]
    \centering
    \begin{subfigure}[b]{0.49\linewidth}
        \centering
        \includegraphics[width=\linewidth]{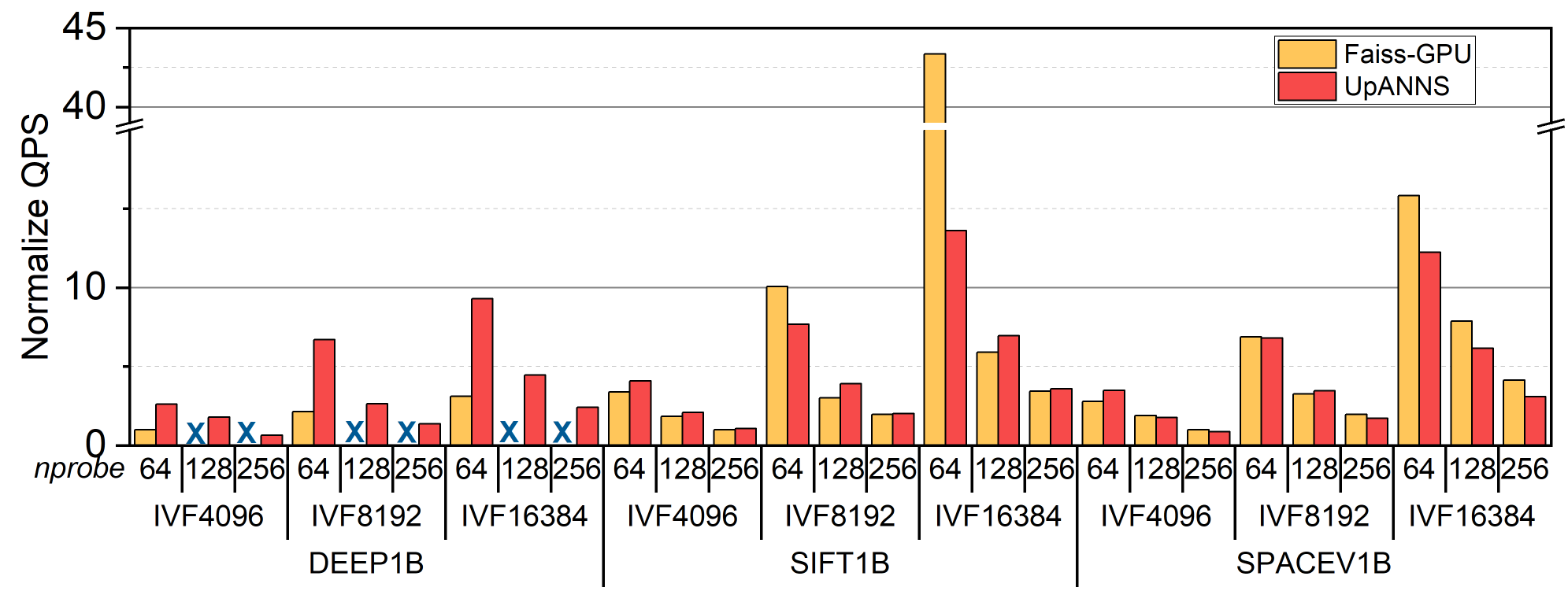}
        \caption{QPS of GPU vs UpANNS}
        \label{fig:overviewgpuqps}
    \end{subfigure}
    \hfil
    \begin{subfigure}[b]{0.49\linewidth}
        \centering
        \includegraphics[width=\linewidth]{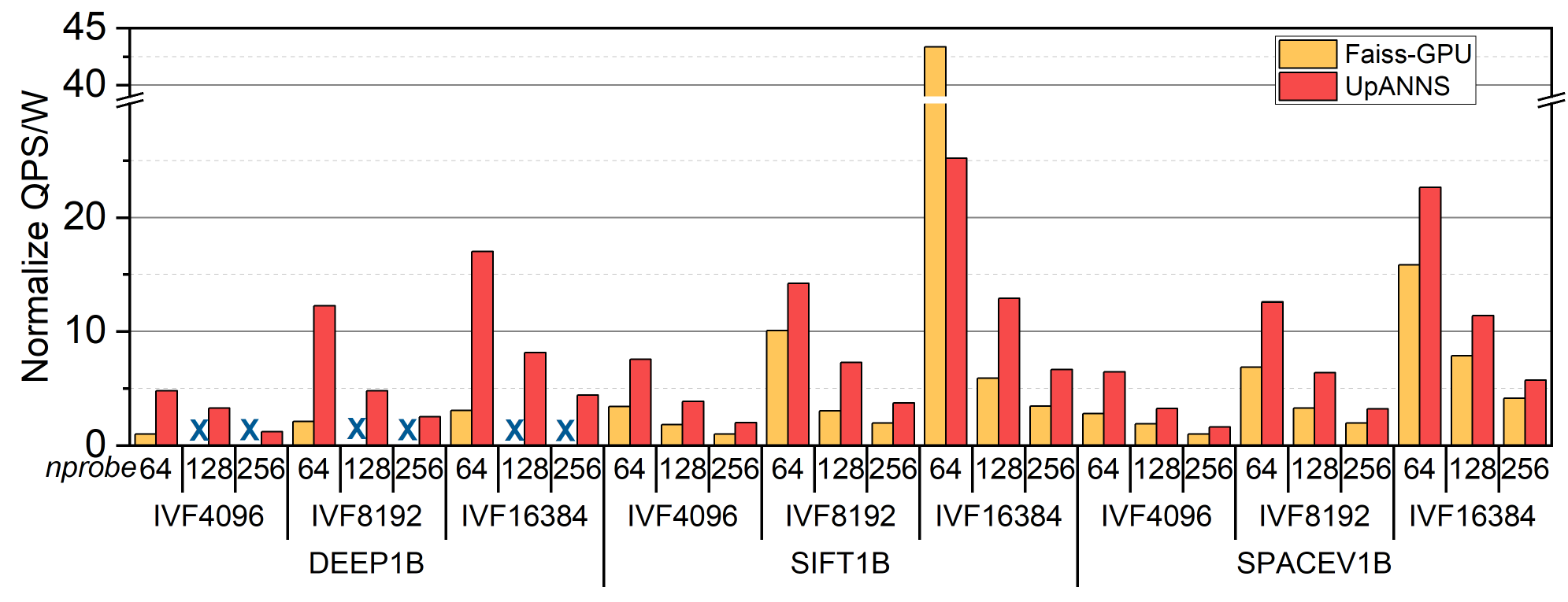}
        \caption{QPS/W of GPU vs. UpANNS}
        \label{fig:overviewgpuqpsperw}
    \end{subfigure}
    \caption{QPS and QPS/W of Faiss-GPU and UpANNS, normalized to those of Faiss-GPU when IVF is 4096 and \emph{nprobe} is 256. \emph{nprobe} varies from 64, 128 to 256. \#clusters varies from 4096, 8192 to 16384. The results are normalized to  in SIFT1B and SPACEV1B, \emph{nprobe} is 64 in DEEP1B.}
    \label{fig:overview}
\end{figure*}

\textbf{Hardware Setup.} 
Table~\ref{tab:setup} shows the hardware specifications of the four compared solutions. Except for the components noted, the remaining hardware is the same across all setups.
For the CPU-based platform, we use two Intel Xeon Silver 4110@2.10GHz CPUs with 4 DDR4-2666Hz DRAM modules, providing 128 GB memory capacity and 85.3 GB/s bandwidth. 
We use one NVIDIA A100 GPU~\cite{gpu} with 80GB memory capacity and 1,935 GB/s bandwidth. For PIM-naive and UpANNS, we use 7 UPMEME PIM modules, with 56 GB total memory capacity and 612.5 GB/s aggregated bandwidth. 
Existing study~\cite{falevoz2023energy} indicates that the peak power of each PIM DIMM is 23.22W, leading to 162W total peak power. 

\textbf{Benchmark.} We evaluate the compared baselines using three billion-scale datasets, namely the DEEP1B~\cite{babenko2016efficient}, SIFT1B~\cite{jegou2011searching} and SPACEV1B~\cite{spacev}. The DEEP1B dataset contains 1 billion 96-dimensional vectors encoded into 12 dimensions. The SIFT1B dataset contains 1 billion 128-dimensional vectors encoded into 16 dimensions. The SPACEV1B dataset contains 1 billion 100-dimensional vectors encoded into 20 dimensions. We process 1,000 queries at a time. The optimizations in UpANNS do not impact the accuracy.

\textbf{Evaluation Metrics.}
We compare different solutions mainly based on Query per second (QPS). For GPU-based Faiss, we also compare the QPS per Watt (QPS/W) to evaluate the cost-effectiveness of different solutions. 
\vspace{2ex}
\subsection{Overall Performance Results}\label{sec::overall-perf}

We first compare the performance and efficiency of the four compared solutions. 
Results are shown in Figure~\ref{fig:overviewpcpu} and Figure~\ref{fig:overview}.
The term IVF4096 indicates that 1 billion points are partitioned into 4096 clusters. We vary \emph{nprobe} from 64, 128 to 256, which indicates the number of clusters selected for each query search. All results are normalized to the values obtained from Faiss-CPU or Faiss-GPU when \emph{nprobe}=256 and IVF=4096 for each dataset. For DEEP1B, due to GPU out-of-memory errors (marked with blue 'X' in Figure~\ref{fig:overview}), results are normalized to Faiss-GPU with \emph{nprobe}=64 and IVF=4096.

Compared to Faiss-CPU and PIM-naive, UpANNS consistently achieves the highest queries per second (QPS) across all settings. Specifically, UpANNS accelerates query search times by 1.6x-3.8x for DEEP1B, by 2.3x-4.3x for SIFT1B and by 2.1x-4.0x for SPACEV1B when compared to Faiss-CPU. Notably, under the same IVF setting, all solutions experience a decrease in QPS as \emph{nprobe} increases, due to the increased search workload. 
Under the same \emph{nprobe}, UpANNS obtains a higher QPS improvement compared to Faiss-CPU when the IVF increases. The reason is that a higher number of clusters results in smaller cluster sizes, which leads to fewer encoded points to search within each cluster and reduces data locality. Hence, the CPU, which has multiple cache layers, does not exhibit a linear increase in QPS with increasing IVF. In contrast, the DPU, which has a small-sized WRAM for cache and MRAM for main memory, is less affected by data locality, enabling UpANNS to achieve greater speedup as IVF increases. 
Although PIM-naive also surpasses Faiss-CPU, it underperforms UpANNS by up to 3.1x, underscoring the critical role of architectural optimizations in the proposed solution.

Comparing UpANNS with Faiss-GPU, Figure~\ref{fig:overviewgpuqps} shows the performance results in terms of QPS, and Figure~\ref{fig:overviewgpuqpsperw} shows the energy efficiency results in terms of QPS per watt. 
Overall, UpANNS obtains comparable QPS as Faiss-GPU in most cases, except when IVF is 16384 and \emph{nprobe} is 64. We used Nvidia Nsight~\cite{nsys} to look into the results and found that the significantly high performance of Faiss-GPU under this setting is because of the different parallelism of the top-k selection. The parallelism of this stage on SIFT1B is 9x higher than the one on SPACEV1B under the same IVF and \emph{nprobe} parameters.
Note that the comparable performance of UpANNS and Faiss-GPU are obtained under a huge gap in the memory bandwidth and computing capabilities between PIM and A100 hardware, as detailed in Table~\ref{tab:setup}. 

From the energy efficiency perspective, the 7 UPMEM DIMMs consume 162W peak power while A100 GPU consumes a significant 300W. Although the actual energy consumption during runtime differs from the peak power, we can use it as an approximation to compare the energy efficiency of UpANNS and Faiss-GPU.
Results in Figure~\ref{fig:overviewgpuqpsperw} show that, UpANNS achieves 2x higher QPS/W compared to Faiss-GPU in most cases, demonstrating better power efficiency. 
While newer GPU models like H100 SXM (~3.5TB/s bandwidth, 700W) and GH200 (~4.9TB/s bandwidth, 1000W) provide 2-3× higher bandwidth, their power requirements scale proportionally, reinforcing UpANNS's position as the more energy-efficient solution.
Beyond energy consumption, we also find that the per dollar QPS of UpANNS can be up to 9.3x higher than that of Faiss-GPU, indicating that UpANNS can significantly reduce costs in real production environments.

\vspace{2ex}
\subsection{Ablation Study}
In this subsection, we conduct ablation studies to evaluate the effectiveness of our key optimizations in UpANNS, including the PIM-Aware Workload Distribution, multi-threading design, Co-occurrence Aware Encoding, and Top-K Pruning strategy.

\subsubsection{PIM-Aware Workload Distribution}
Since the workloads need to be distributed among DPUs and executed in parallel, the largest workload among DPUs determines the overall performance. We evaluate the performance improvement due to the PIM-Aware Workload Distribution strategy compared to the naive distribution strategy that assigns clusters randomly to DPUs. Figure~\ref{fig:scheduled-try} shows the ratio of maximum process to average process under different \emph{nprobe} and IVF settings. A ratio closer to 1 indicates that the workload is more evenly distributed among DPUs. We observe that the ratio for PIM-naive is significantly higher than 1, especially when the IVF and \emph{nprobe} are small. In contrast, the ratio for UpANNS is close to 1 under all settings for all datasets, indicating that the PIM-Aware Workload Distribution strategy effectively balances the workload among DPUs.
\vspace{2ex}
\subsubsection{Number of Threads per DPU}

\begin{figure}[t]
    \centering
    \includegraphics[width=1\linewidth]{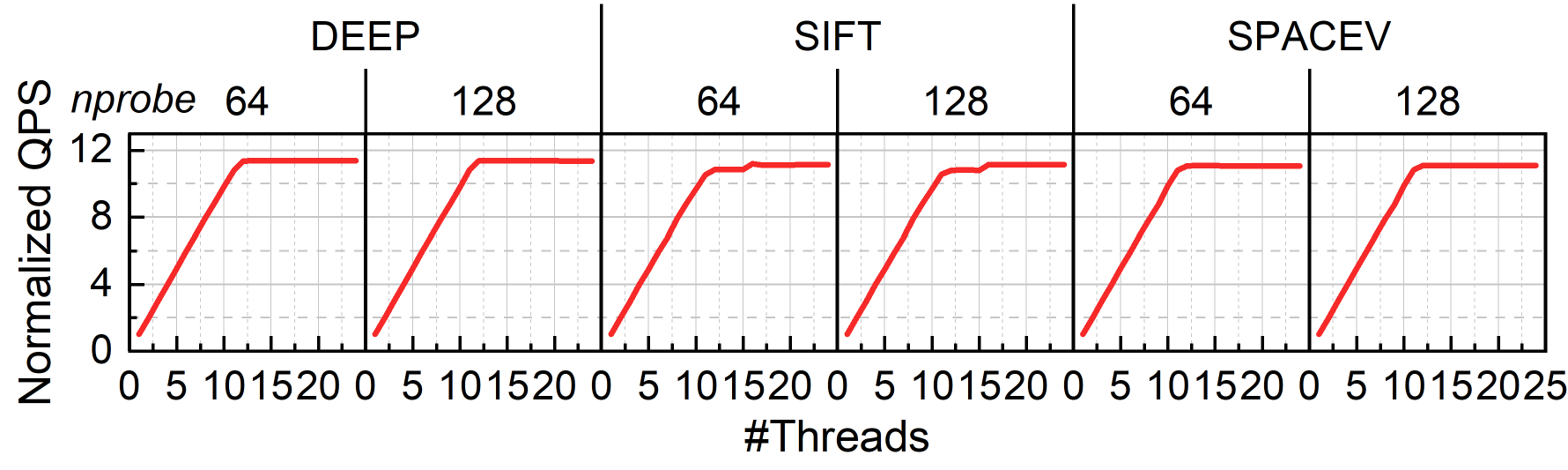}
    \caption{QPS of UpANNS as the \#threads increases}
    \label{fig:tasklets-try}
\end{figure}
UPMEM PIM could support up to 24 threads per DPU~\cite{upmemstat}. Therefore, we evaluate the performance of UpANNS by varying the number of threads from 1 to 24. Figure~\ref{fig:tasklets-try} shows our results for three datasets with different \emph{nprobe} values. We normalize the values to those when the number of tasklets is one. 
We have similar observations in all settings. That is, the QPS increases linearly as the number of tasklets increases up to 11. Beyond 11 tasklets, the performance nearly saturates. The QPS of UpANNS with 11 tasklets is almost 11x higher than that with a single tasklet. 
This is because each DPU has a 14-stage pipeline, and only the last three stages can execute in parallel with the first two stages of the next instruction within the same thread. Using more than 11 tasklets makes full use of the pipeline and keeps the DPU busy. Thus, by default, we set \#threads to 11 per DPU.
\vspace{2ex}
\subsubsection{Co-occurrence Aware Encoding}

\begin{figure}[t]
    \centering
        \hspace*{-1.5em}
    \includegraphics[width=1.1\linewidth]{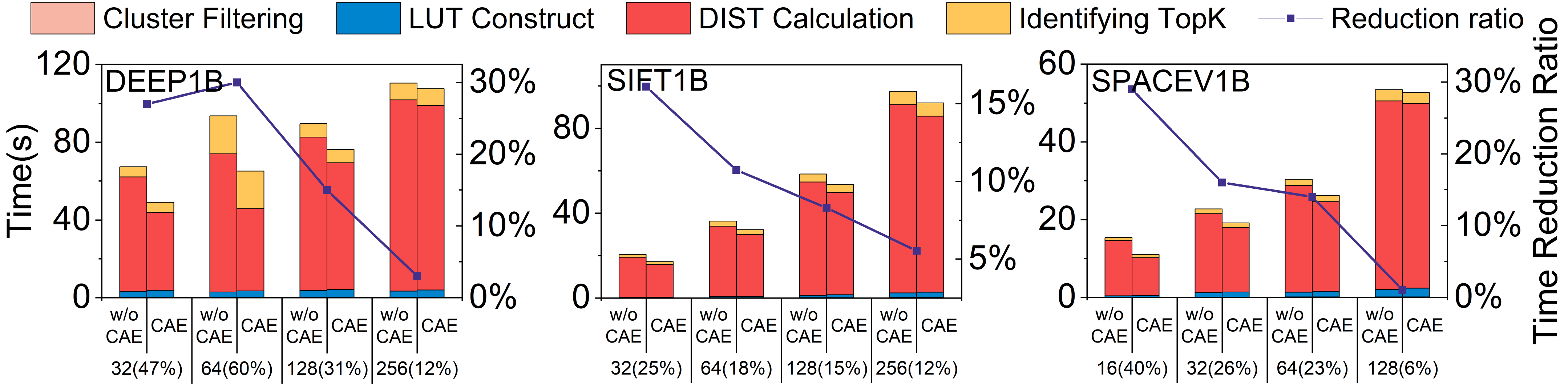}
    \caption{Performance improvement due to Co-occurrence Aware Encode strategy (CAE) under different \emph{nprobe} and corresponding length reduction rates.}
    \label{fig:cache-exp}
\end{figure}

To evaluate the performance improvement due to the Co-occurrence Aware Encoding strategy, we select queries that choose top-\emph{nprobe} clusters with the highest vector length reduction rates and calculate the average vector length reduction rate in the maximum workload DPU. Figure~\ref{fig:cache-exp} shows the performance improvement due to the Co-occurrence Aware Encoding strategy under different \emph{nprobe} values and corresponding length reduction rates. We observe the following for all settings. First, the performance improvement correlates positively with the length reduction rate, which is consistent with our design. 
Second, the LUT construction time increases slightly due to the additional time needed to construct the partial sum.
Third, the performance improvement becomes more significant with higher length reduction rates. This is because a higher length reduction rate means less data needs to be transferred between MRAM and WRAM, allowing calculations to be done more efficiently. The time breakdown also shows that distance calculation time decreases more when the length reduction rate is higher.
\vspace{2ex}
\subsubsection{Top-K Pruning}
\begin{figure}[t]
    \centering
    \begin{minipage}{.23\textwidth}
    \hspace*{-1em}
        \includegraphics[width=1\linewidth]{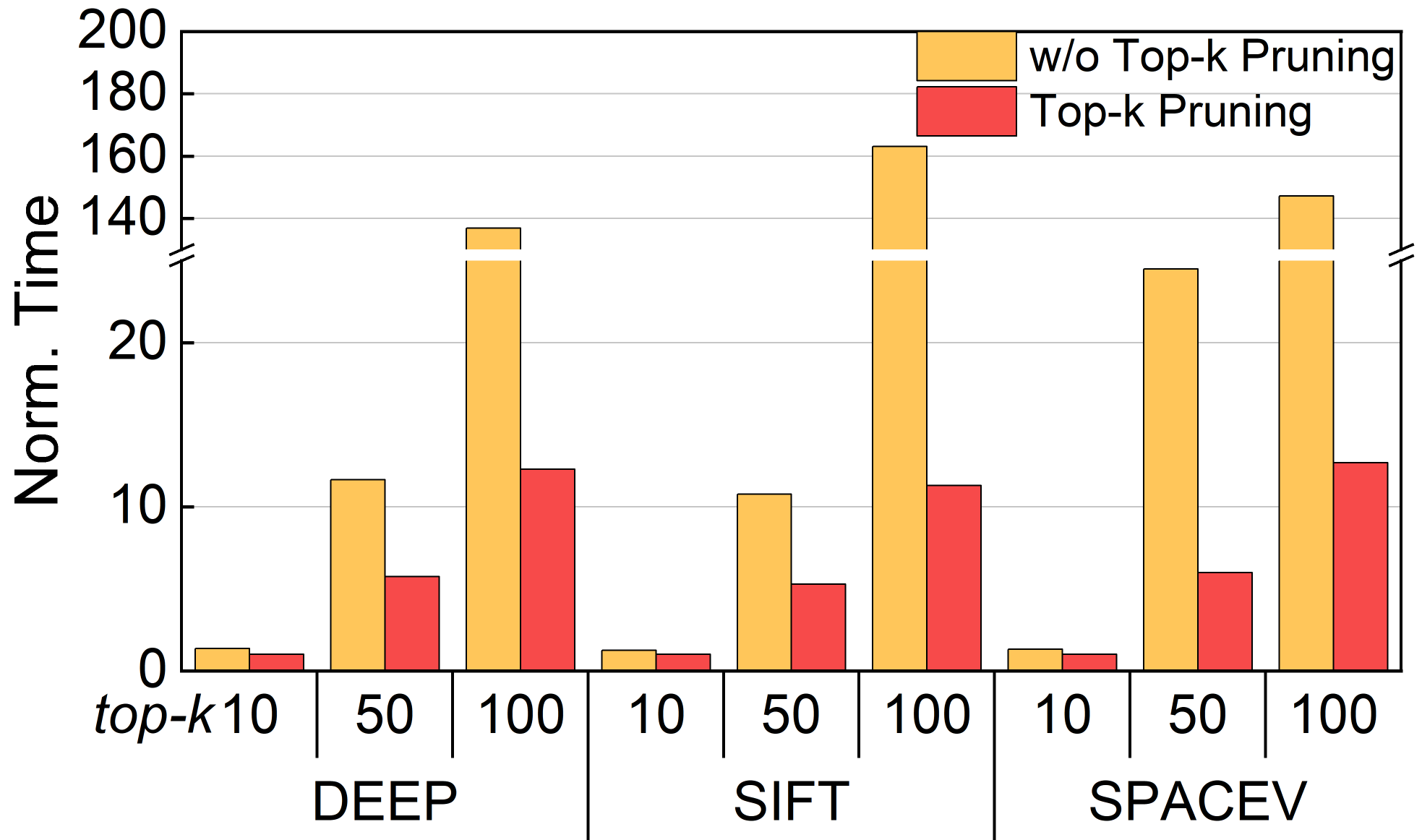}
    \caption{Time reduction with proposed top-k selection strategy.}
    \label{fig:topk-ablation}
    \end{minipage}
    \hfill
    \begin{minipage}{.23\textwidth}
        \includegraphics[width=1\linewidth]{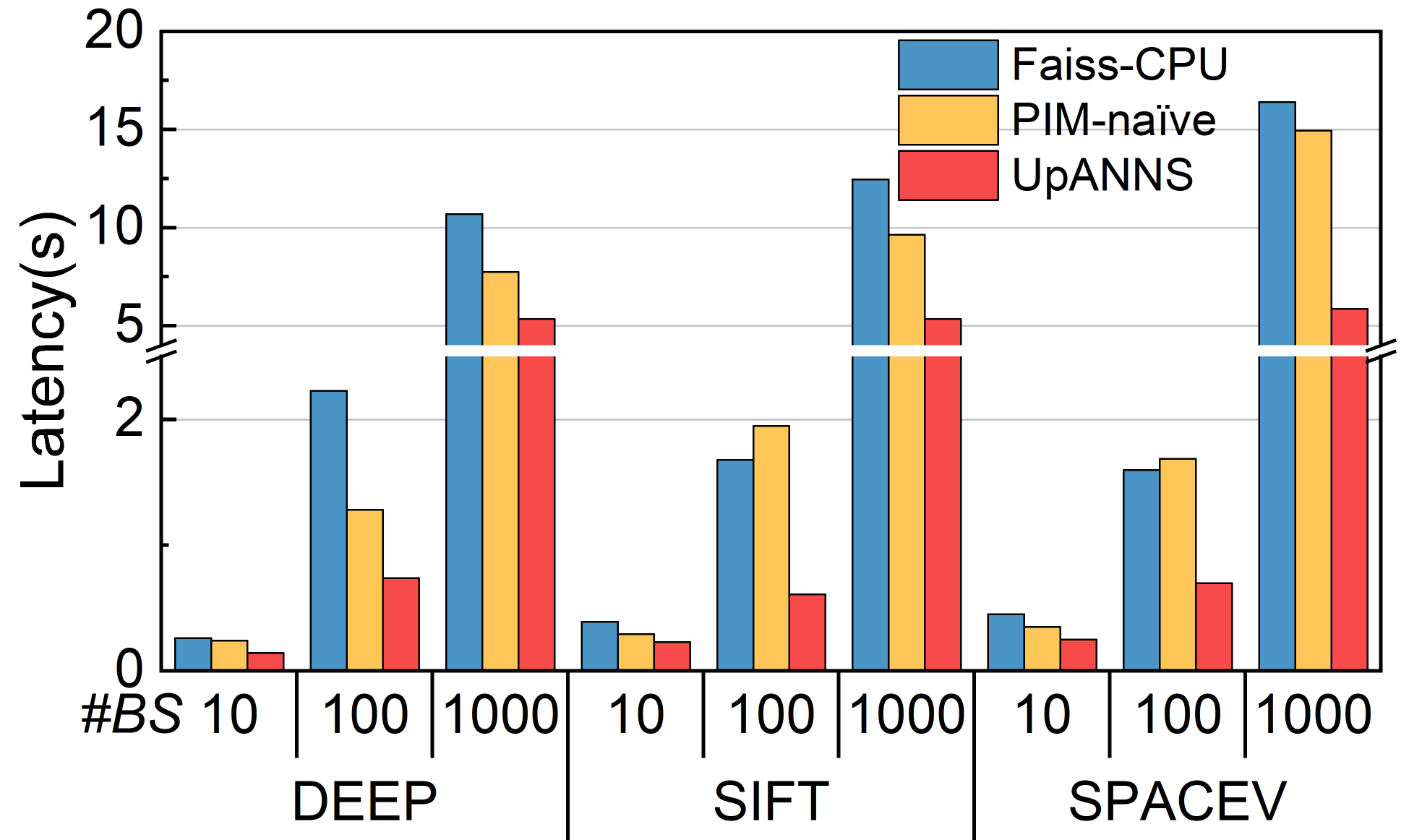}
    \caption{The impact of different batch size on query latency.}
    \label{fig:bs-sensitivity}
    \end{minipage}
\end{figure}

We evaluate the time of the top-k selection stage with and without the proposed top-k selection strategy. Figure~\ref{fig:topk-ablation} shows the time reduction due to our proposed strategy. The results are normalized to the time with the proposed strategy in top-10. We observe that the top-k selection time increases linearly with the top-k value. The proposed strategy significantly reduces the time for top-k selection, especially when the top-k value is large. This is because our strategy reduces the number of comparisons needed to find the top-k results.


\vspace{2ex}
\subsection{Sensitivity Study}
We study the impact of three parameters to the performance of UpANNS, including the batch size of queries,
the data size of each MRAM read and the required k size in the top-k query.
\vspace{2ex}
\subsubsection{Batch Size}
Figure~\ref{fig:bs-sensitivity} shows the impact of different batch sizes on query latency. We set the IVF to 4096 and \emph{nprobe} to 64, varying the batch size (\#BS) from 10, 100 to 1000. We observe that UpANNS achieves the lowest query latency in all settings, and the speedup of UpANNS over Faiss-CPU and PIM-naive increases as the batch size increases. This is because the overhead of the pre-processing and post-processing stages is amortized over a larger number of queries. This makes it suitable for real-world applications where the number of queries is usually large.

\vspace{2ex}
\subsubsection{MRAM Read Size}

As described in Section~\ref{sec::multi-threading}, the MRAM read latency does not increase linearly with size. To decide an optimal MRAM read size for each thread, we evaluate the performance of UpANNS with varied MRAM read sizes. Since the MRAM read size must be between 8 Bytes - 2048 Bytes and aligned with 8 Bytes, we varied the number of vectors read at once. For example, we vary the number of vectors fetched in one MRAM read from 2, 4, ..., to 64, making the MRAM read size vary from 64 Bytes, 128 Bytes, ..., to 2 KB to fit in the 2048 Bytes limit in SIFT1B.
We set IVF to 4096 and top-k to 10. Figure~\ref{fig:readsize} shows the evaluation results.

For all datasets, the QPS increases more rapidly when the number of vectors increases from 2 to 16, while becoming much more stable when the number of vectors exceeds 16. This is consistent with our observation in Figure~\ref{fig:mram:latency}, which shows that the MRAM read latency grows slowly when the data transfer size increases from 8 Bytes to 512 Bytes and grows dramatically beyond 512 Bytes. 
By default, we set the MRAM read size to 16 vectors to have good QPS and reasonable WRAM size at the same time.


\subsubsection{Top-K Size}\label{sec::sensitivity}



\begin{figure}[t]
    \centering
    \begin{minipage}{.23\textwidth}
    \hspace*{-1em}
        \includegraphics[width=1\linewidth]{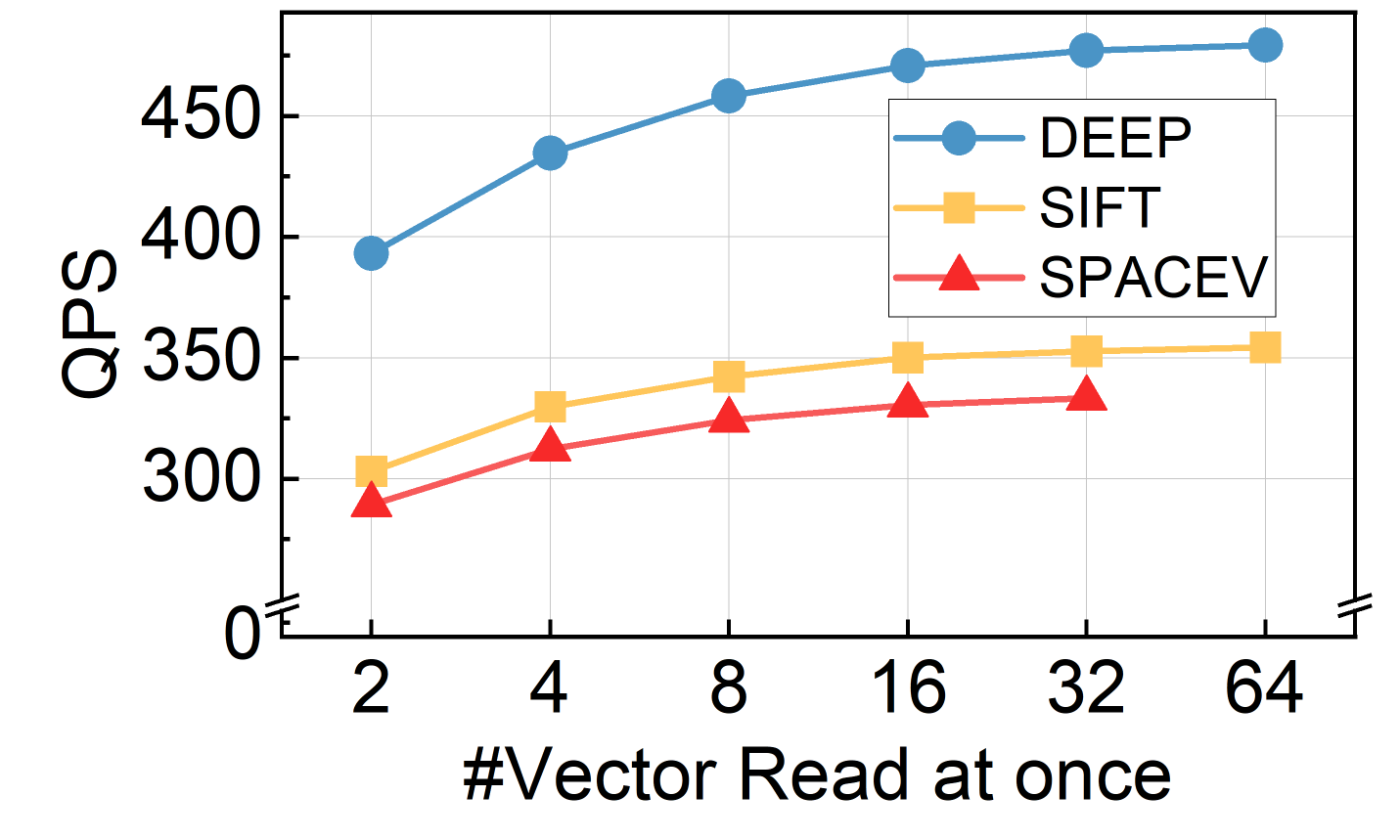}
    \caption{The impact of MRAM read size.}
    \label{fig:readsize}
    \end{minipage}
    \hfill
    \begin{minipage}{.23\textwidth}
        \includegraphics[width=1\linewidth]{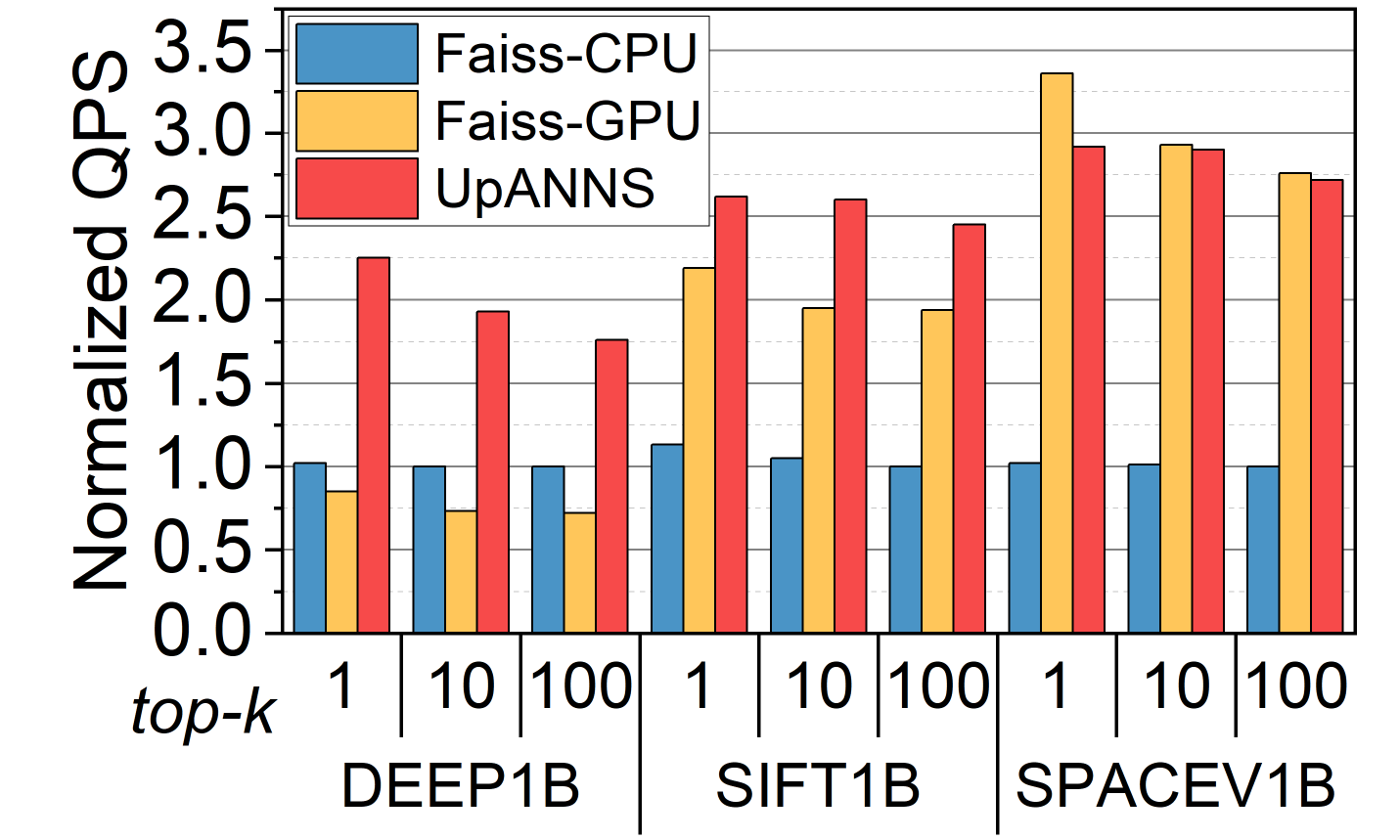}
    \caption{The impact of top-k size.}
    \label{fig:topk}
    \end{minipage}
\end{figure}

\begin{figure}[t]
    \centering
    \includegraphics[width=1\linewidth]{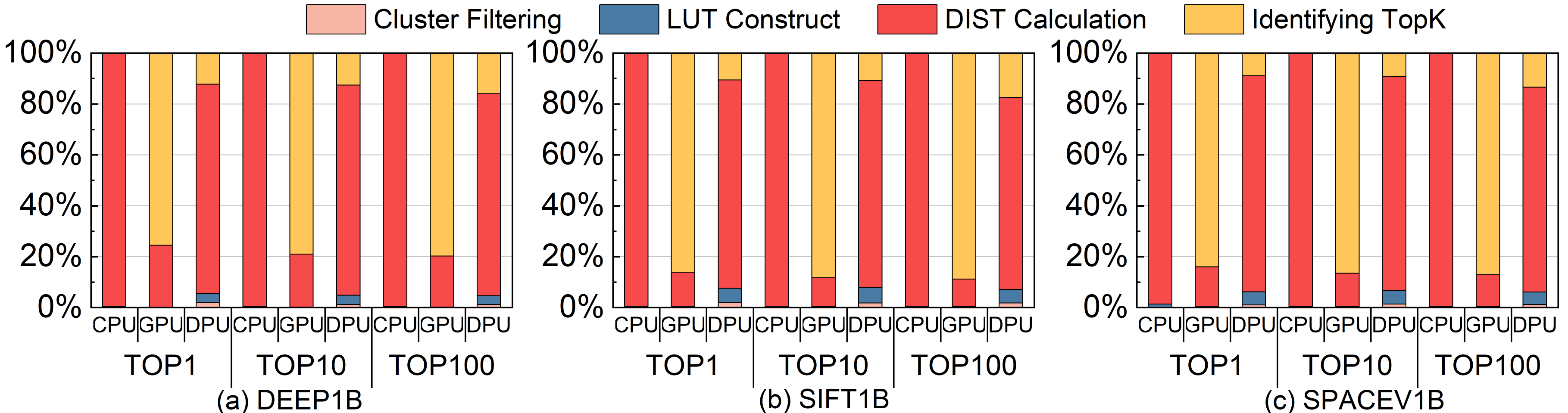}
    \caption{Query search time breakdown}
    \label{fig:breakdown-exp}
\end{figure}


To evaluate the impact of top-k size, we vary k from 1 to 100 while maintaining default parameters. Figure~\ref{fig:topk} presents normalized QPS results (relative to Faiss-CPU’s top-100 performance) for UpANNS, Faiss-CPU, and Faiss-GPU. UpANNS achieves consistent performance advantages, achieving an average 2.5x higher QPS than Faiss-CPU and 1.6x higher than Faiss-GPU across datasets. While Faiss-CPU’s QPS remains stable across k values, UpANNS and Faiss-GPU exhibit slight QPS degradation as k increases. This stems from the growing top-k list size, which amplifies CPU-DPU communication costs for UpANNS and CUDA stream synchronization overhead for Faiss-GPU during top-k selection.

A detailed processing time breakdown (Figure~\ref{fig:breakdown-exp}) reveals critical insights. UpANNS significantly reduces the distance calculation stage’s time ratio from 99.5\% (Faiss-CPU) to 75.5–80\% across datasets, demonstrating its effectiveness in mitigating CPU memory bottlenecks. While GPUs leverage high memory bandwidth to alleviate memory access challenges, their performance is constrained by CUDA synchronization during identifying top-k, consuming over 85\% of processing time for large datasets. DPUs avoid this bottleneck, achieving GPU-competitive performance. As k increases, the top-k stage’s time ratio grows from 76\% to 89\% on GPUs and 9\% to 17\% on DPUs, while remaining negligible on CPUs due to their dominance by distance calculations. These trends align with the observed QPS variations under different k sizes.
\vspace{2ex}
\subsection{Scalability Study}


\begin{figure}[t]
    \Description{This figure shows the normalized QPS of different solutions with varying nprobe and cluster counts.}
    \centering
    \includegraphics[width=\linewidth]{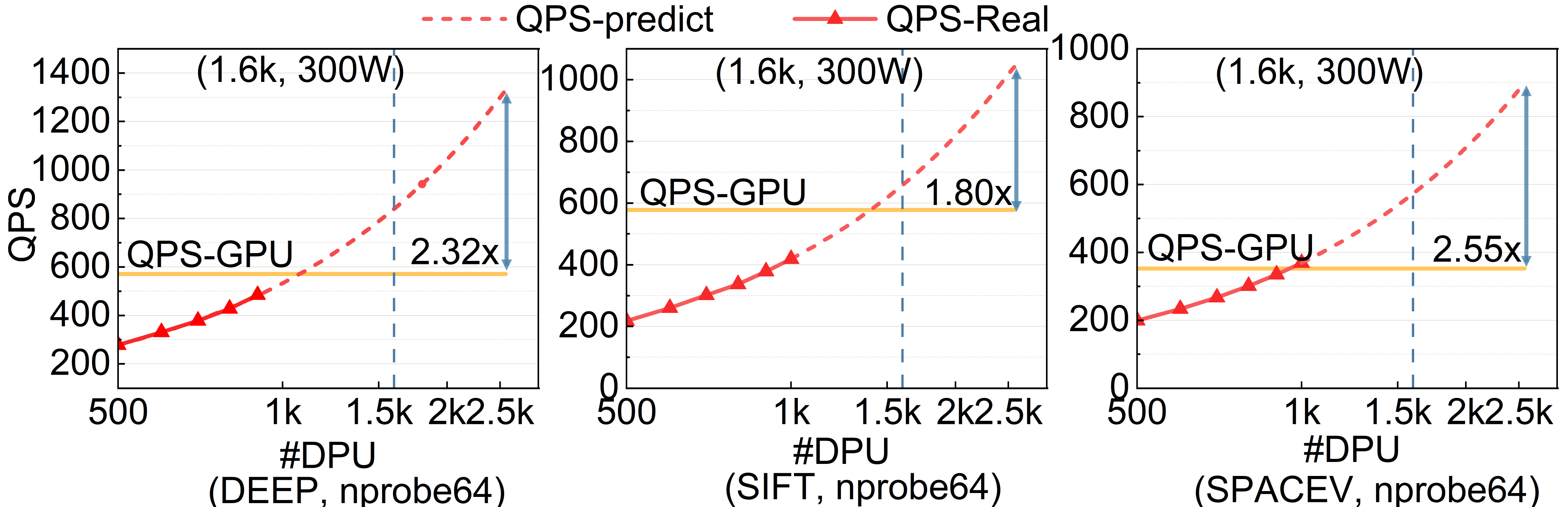}
    \caption{Scalability of UpANNS under different \#DPUs. Red triangles are measured from real hardware and red dash line represents predicted QPS. The yellow line indicates the QPS of Faiss-GPU. The blue vertical dash line shows when DPUs consume power equal to that of an A100 GPU.}
    \label{fig:scale}
\end{figure}

Existing systems can hold up to 20 UPMEM DIMMs, each containing 128 DPUs~\cite{gomez2021benchmarking}, totaling 2560 DPUs. 
To study the performance of UpANNS beyond the seven PIM DIMMs we have, we evaluate the QPS of UpANNS with 500, 600, ..., 900 DPUs and use the regression method to predict the QPS of UpANNS with up to 2560 DPUs. Due to the limited memory capacity when the number of DPUs is low, we evaluate the performance using 500 million scale.

As shown in Figure~\ref{fig:scale}, the regression curve fits perfectly with the QPS results measured from 500-900 DPUs. When the number of DPUs increases, the QPS of UpANNS increases almost linearly, demonstrating good scalability of our system.
When the number of DPUs is 2560, UpANNS can achieve up to 2.6x higher QPS compared to Faiss-GPU. Note that, even with 20 DIMMs, the cost of PIM is still much lower than that of A100 GPU (\$8000 vs. \$20,000).
If we compare the performance of UpANNS and Faiss-GPU under the same peak power (i.e., 300W), we can use 1654 DPUs. As shown by the blue vertical dash line in Figure~\ref{fig:scale}, UpANNS obtains higher QPS than Faiss-GPU at the same peak power constraint under all settings.
This again demonstrates the energy efficiency of UpANNS, making it practical for real-world large-scale production systems.
While our current evaluation focuses on single-host configurations, UpANNS can be easily extended to multi-host configurations. Only query distribution and result aggregation require cross-host communication. The core memory-intensive search operations remain local to each host, ensuring efficient scalability.


\vspace{2ex}
\section{Related Work}\label{sec::related}


Existing methods to improve ANNS performance can be categorized into {software-based} and {hardware-based methods}.


\textbf{Software-based Methods.} Software-based methods primarily enhance the performance of ANNS through algorithm-level optimizations. For instance, VDTuner~\cite{vdtuner} utilizes a learning-based parameter tuning method to more quickly find the optimal parameter configuration for ANNS databases. Works such as ~\cite{andre2016cache,adaptive,gao2023high} reduce memory access frequency by establishing a distance bound or using pruning methods. In contrast, our proposed method accelerates the ANNS algorithm based on new hardware, complementing the aforementioned software-based methods.


\textbf{Hardware-based Methods.} Hardware-based methods primarily accelerate ANNS by utilizing existing hardware or designing new hardware. GPUs are the most commonly used hardware to accelerate ANNS. For example, works such as~\cite{zhao2020song,khan2024bang,groh2022ggnn} employ GPU-optimized graph algorithms for ANNS but face scalability challenges due to limited GPU memory. Juno~\cite{juno-asplos24} addresses this using GPU ray-tracing cores (RT-cores) to accelerate the IVFPQ compression-based method, reducing storage costs compared to graph-based methods. However, GPU solutions generally exhibit lower energy efficiency than UpANNS.


In addition to GPUs, many other hardware options have also been used to accelerate ANNS. For example, works such as~\cite{abdelhadi2019accelerated,vectorsearch-sc23, yuan2025fanns} utilize FPGA to achieve high throughput on million-scale datasets but struggle with billion-scale data due to constrained on-chip memory. SmartSSD-based systems~\cite{9726805,haikun-atc24} offer ample memory but risk PCIe bandwidth bottlenecks when coexisting with applications like LLMs. Compute Express Link (CXL)~\cite{cxl-anns} accelerates ANNS by expanding memory via additional controllers, yet raises system maintenance costs. In contrast, UpANNS is a cost-effective and easily scalable solution that offers near-linear performance scalability for large-scale datasets.

In addition to existing hardware, many works design custom hardware specifically tailored to the characteristics of ANNS for acceleration. For example, \cite{lee2022anna} proposed a specialized architecture, which combines the benefits of a specialized dataflow pipeline and efficient data reuse to accelerate ANNS. The work~\cite{wang2024ndsearch} proposed a near-data processing (NDP) architecture based on SmartSSD to accelerate the graph-traversal-based ANNS task. However, the aforementioned works are still at the simulation or prototype stage, and are far from practical application.  


\section{Conclusion}
In this work, we addressed the significant performance bottlenecks faced by CPU and GPU-based Approximate Nearest Neighbor Search (ANNS) solutions at billion-scale datasets, where CPU-based solutions are bounded by limited memory bandwidth and GPU-based solutions encounter memory capacity and resource utilization issues. We present UpANNS, a novel framework leveraging UPMEM's Processing-in-Memory (PIM) architecture to overcome the memory bottleneck in billion-scale
ANNS algorithms. UpANNS effectively addresses memory bottlenecks by employing architecture-aware data placement, efficient resource management, and a novel encoding approach for the IVFPQ algorithm. Our extensive evaluation demonstrates that UpANNS significantly improves performance, achieving a remarkable 4.3x increase in QPS compared to CPU-based implementations of Faiss, while also matching the performance of GPU-based Faiss systems. Furthermore, UpANNS shows a commendable 2.3x improvement in QPS per Watt compared to GPU solutions, highlighting its superior cost-effectiveness and potential for large-scale applications such as large model serving.

Although our current implementation is tuned for IVFPQ, the core techniques, namely workload distribution, resource management, and top-k pruning, are transferable. Future work will generalize UpANNS to broader ANNS algorithms, and exploit next-generation PIM hardware with higher frequency and bandwidth to further improve competitiveness against high-end accelerators.


\bibliographystyle{ACM-Reference-Format}
\bibliography{refs}

\end{document}